\documentclass[10pt,final,doublecolumn]{IEEEtran}

\usepackage[caption=false,font=normalsize,labelfont=sf,textfont=sf]{subfig}
\usepackage{amsmath}
\usepackage{amssymb}   

\usepackage{titlesec}
\titlespacing*{\subsection}{0pt}{0.5\baselineskip}{0.3\baselineskip}
\usepackage{amsthm}
\usepackage{graphicx}
\usepackage{bbding}
\usepackage{indentfirst}
\usepackage{algorithm}
\usepackage{setspace}
\usepackage{float}
\usepackage{amsfonts}
\usepackage{enumerate}
\usepackage{multicol}
\usepackage{color}
\usepackage{bm}
\usepackage{stfloats}
\usepackage{epstopdf}
\usepackage{threeparttable}
\usepackage{makecell}
\usepackage{cite}

\usepackage{booktabs}

\usepackage{algorithm}
\usepackage{algorithmic}

\newcommand{\mbf}[1]{\mathbf{#1}}
\newcommand{\mbb}[1]{\mathbb{#1}}
\newcommand{\mc}[1]{\mathcal{#1}}

\theoremstyle{remark}

\begin{document}

	\title{Multi-Cell 6DMA: Cooperative Interference
		Management and Antenna Rotation Optimization}
	
	\author{
		Qijun Jiang,~\IEEEmembership{Graduate Student Member,~IEEE}, 
		Xiaodan Shao,~\IEEEmembership{Member,~IEEE}, \\
		and Rui Zhang,~\IEEEmembership{Fellow,~IEEE}

	\thanks{Q. Jiang is with the School of Science and Engineering, The Chinese University of Hong Kong, Shenzhen, Guangdong 518172, China (e-mail: qijunjiang@link.cuhk.edu.cn).}%
	\thanks{X. Shao is with the Department of Electrical and Computer Engineering, University of Waterloo, Waterloo, ON N2L 3G1, Canada (e-mail: x6shao@uwaterloo.ca).}%
	\thanks{R. Zhang is with the Department of Electrical and Computer Engineering, National University of Singapore, Singapore 117583 (e-mail: elezhang@nus.edu.sg).}%

}
\maketitle

\begin{abstract}
	In this paper, we investigate a multi-cell six-dimensional movable antenna (6DMA) network for enhancing downlink communication performance under inter-cell interference (ICI). Each base station (BS) is equipped with multiple 6DMA surfaces, and the 6DMA rotations affect both the desired-signal enhancement for in-cell users and the interference leakage toward neighboring cells, which makes the antenna-rotation design and transmit precoding intrinsically coupled across BSs. To address this issue, we formulate an average weighted sum-rate maximization problem for the multi-cell system by jointly optimizing the short-term downlink precoders and long-term 6DMA rotations under practical antenna geometric constraints. To tackle the resulting nonconvex problem, we propose a distributed two-timescale design based on inter-cell interference power constraint (IPC) coordination among neighboring BSs, under which each BS performs local short-term precoder optimization based on instantaneous channel state information (CSI) and long-term 6DMA rotation update according to statistical CSI with limited inter-BS information exchange. In particular, an edge-wise IPC coordination mechanism based on two-stage one-dimensional grid search and random maximal matching is developed to enable scalable distributed implementation. A centralized offline benchmark is also provided for performance comparison. Numerical results show that the proposed   distributed  design achieves performance close to the centralized benchmark under different interference conditions, while maintaining favorable scalability as the network size increases.
\end{abstract}


\section{Introduction}
Multi-antenna transmission has been one of the key driving forces behind the evolution of wireless networks. By exploiting the spatial degrees of freedom of wireless channels, multiple-input multiple-output (MIMO) systems can achieve substantial multiplexing and diversity gains, thereby significantly improving the throughput and reliability of wireless communications \cite{MIMO,10054381}. To further enhance the utilization of the wireless spatial domain, a variety of emerging technologies have been investigated, such as intelligent reflecting surfaces (IRSs) \cite{9133130,10555049,renwang1,10304548}, coordinated multi-point (CoMP) \cite{CoMP}, cell-free massive MIMO \cite{Cell_free}, and fluid/movable antennas\cite{FAS, zhu2024modeling,zhu2024movable}. Among them, fluid/movable antennas improve channel conditions by flexibly adjusting antenna positions at the wavelength scale within a given spatial region. However, the performance gain brought by wavelength-scale position variation with fixed antenna orientation is still limited. To more effectively exploit spatial channel variations, the recently proposed six-dimensional movable antenna (6DMA) architecture further enriches the wireless design space by enabling the adaptation of the three-dimensional (3D) positions and/or 3D rotations of antennas or antenna sub-surfaces, thus providing substantially greater flexibility for reshaping wireless channels than conventional fixed-position arrays \cite{6DMA_TWC,6dma_dis,6DMA_Qijun,10945745, 11148174}. Following the initial 6DMA framework, recent studies have extended this research direction to several important scenarios, including  6DMA channel estimation by exploiting a new directional sparsity \cite{6DMA_JSTSP}, 6DMA-enhanced IRS systems \cite{qingmma}, 6DMA for the Internet of Vehicles \cite{ji20266d,ji2026single}, 6DMA-enabled wideband terahertz (THz) communications \cite{yan2025six}, polarlized 6DMA \cite{IPA}, six-dimensional movable holographic-surface-based integrated data and energy transfer systems \cite{wang20256d}, antenna rotation optimization \cite{6DMA_Xiaodan_2}, flexible-sector 6DMA \cite{6DMA_flexible_sector}, hybrid near-far-field 6DMA \cite{near}, channel knowledge map-aided 6DMA \cite{11314850}, unmanned aerial vehicle (UAV)-enabled passive 6DMA \cite{liu2024uav}, flexible coupler antennas (FCAs) with translatable/rotatable couplers by exploiting controllable mutual coupling \cite{86240082,FC1,FCjstsp,FCATWC,RCATWC}, and antenna position optimization \cite{li2025ai}. These advances have demonstrated the strong potential of 6DMA for future wireless systems with enhanced spatial reconfigurability.

Despite these promising developments, existing 6DMA studies have mainly focused on point-to-point communications, single-cell systems, centralized multiuser settings, or other specialized architectures \cite{6DMA_TWC,6DMA_Xiaodan_2,6DMA_Qijun}. However, the exploitation of 6DMA in interference-limited multi-cell wireless networks with distributed base station (BS) deployment remains largely unexplored. In such systems, 6DMA rotations affect both the desired-signal enhancement for in-cell users and the interference leakage toward neighboring cells, which intrinsically couples the antenna-rotation design and transmit precoding across different BSs. Moreover, the resulting design is inherently two-timescale, where the short-term precoders adapt to instantaneous channel state information (CSI), while the long-term 6DMA rotations are optimized according to user distributions and channel statistics. A fully centralized solution would require global instantaneous CSI and network-wide statistical information, which leads to excessive backhaul signaling and rapidly increasing computational complexity as the network size grows.

Meanwhile, distributed coordination has been extensively studied for conventional multi-cell systems with fixed-position BS arrays, including coordinated beamforming, interference pricing, and cooperative precoding \cite{Coordinatedbeamforming,Interferencepricing,Cooperativeprecoding}. However, these methods are not directly applicable to 6DMA-enabled multi-cell networks, because 6DMA introduces additional long-term spatial control variables (antenna rotations)  subject to practical geometric constraints, which fundamentally reshape the channel responses and interference conditions. As a result, independently optimizing the 6DMA rotations at each BS within existing distributed beamforming frameworks may incur substantial performance loss, whereas a joint optimization over all BSs is difficult to scale. Although decentralized or cell-free 6DMA designs have recently emerged \cite{6DMA_cellfree}, their architectures and coordination mechanisms differ significantly from those in practical multi-cell cellular networks, where each BS mainly serves its own users and exchanges only limited information with neighboring BSs. Therefore, a new framework is needed for multi-cell 6DMA systems to jointly address inter-cell interference (ICI) management, long-term rotation design, and scalable distributed short-term precoder optimization.

Motivated by the above observations, this paper studies a downlink multi-cell 6DMA network, where each BS is equipped with multiple 6DMA surfaces mounted on a circular track and each surface can adjust its azimuth and tilt angles. We aim to maximize the average weighted sum-rate by jointly optimizing the short-term downlink precoders and long-term 6DMA rotations. To this end, we propose a distributed two-timescale design with limited inter-BS coordination, in which inter-cell interference power constraint (IPC) thresholds serve as a low-dimensional coordination interface among neighboring BSs. In addition, we establish a centralized offline benchmark as a performance reference for evaluating the proposed distributed design.
The main contributions of this paper are summarized as follows.
\begin{itemize}
	\item We propose a multi-cell 6DMA design framework and formulate a stochastic two-timescale optimization problem for average weighted sum-rate maximization. The formulation captures the coupling between short-term precoder adaptation and long-term 6DMA rotation design under practical geometric constraints on the antenna surfaces.
	
	\item We develop an IPC-assisted distributed short-term precoder design for multi-cell 6DMA systems. Under coordinated IPC thresholds, each BS performs local precoder optimization without requiring exact instantaneous interference information from other cells, thereby significantly reducing inter-BS signaling overhead.
	
\item We further develop a distributed long-term design that combines edge-wise IPC coordination with per-BS 6DMA rotation optimization. In particular, an edge-wise IPC coordination mechanism based on local rate evaluation, two-stage one-dimensional grid search, and random maximal matching is proposed to enable scalable distributed implementation.
	
	\item We analyze the computational complexity of the proposed algorithm and show that its per-BS complexity does not explicitly scale with the number of cells. Numerical results show that the proposed method achieves performance close to the centralized benchmark under different interference levels and remains effective as the network size increases.
\end{itemize}
\begin{figure}[t]
	\centering
	\includegraphics[width=0.6\linewidth]{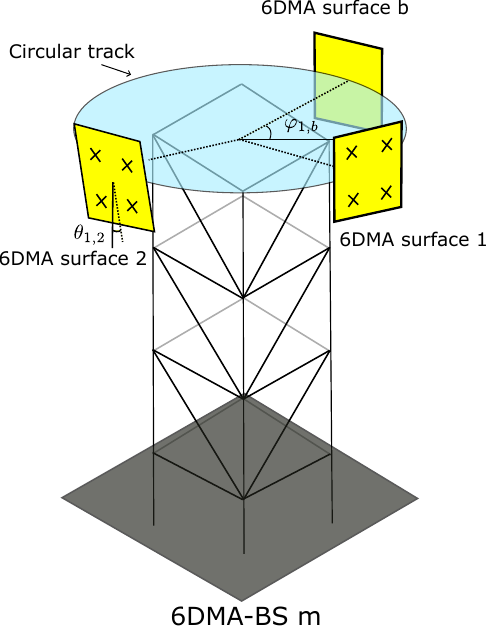} 
	\caption{Illustration of the 6DMA-BS architecture. $B$ 6DMA surfaces are mounted on a circular track at BS $m$, where each surface $b$ can adjust its azimuth direction $\varphi_{m,b}$  and  tilt angle $\theta_{m,b}$.}

	\label{fig:circular_model}
	\end{figure}
The rest of this paper is organized as follows. Section II presents the multi-cell 6DMA system model and formulates the two-timescale optimization problem. Section III develops the centralized offline benchmark. Section IV proposes the distributed IPC-assisted two-timescale algorithm. Section V provides simulation results. Section VI concludes this paper.

\textit{Notations:} Boldface lowercase and uppercase letters denote vectors and matrices, respectively, while calligraphic letters denote sets. For a scalar, vector, or matrix, $(\cdot)^*$, $(\cdot)^T$, and $(\cdot)^H$ denote the conjugate, transpose, and conjugate transpose, respectively. $\mathbb{R}$, $\mathbb{C}$, and $\mathbb{R}_+$ denote the sets of real numbers, complex numbers, and nonnegative real numbers, respectively. $\mathcal{CN}(0,\cdot)$ denotes the circularly symmetric complex Gaussian distribution. $\mathbb{E}\{\cdot\}$ denotes expectation, $|\cdot|$ denotes the absolute value of a scalar or the cardinality of a set, $\|\cdot\|_2$ denotes the Euclidean norm, and $\Re\{\cdot\}$ denotes the real part of a complex number. $\mathbf{I}$ and $\mathbf{0}$ denote the identity matrix and the all-zero vector or matrix with appropriate dimensions, respectively. For a vector $\mathbf{a}$, $[\mathbf{a}]_i$ denotes its $i$-th entry, while for a matrix $\mathbf{A}$, $[\mathbf{A}]_{i,j}$ denotes its $(i,j)$-th entry.  $\mathcal{O}(\cdot)$ denotes the big-O complexity notation.

\section{System Model and Problem Formulation}

\subsection{Channel Model}
\label{subsec:SystemModel}

We consider a downlink cellular network with $M$ six-dimensional movable antenna (6DMA)-equipped base stations (6DMA-BSs), each serving one cell, and denote the corresponding BS index set by $\mc{M}\triangleq\{1,\ldots,M\}$.
Each 6DMA-BS is equipped with $B$ 6DMA surfaces, indexed by $\mc{B}\triangleq\{1,\ldots,B\}$. Each surface contains $N$ antennas, indexed by $\mc{N}\triangleq\{1,\ldots,N\}$. The fixed position of the $m$-th 6DMA-BS is denoted by $\mathbf{o}_m\in\mathbb{R}^{3\times 1}$.
Unlike conventional antenna arrays deployed at fixed positions (e.g., sectored antenna arrays), the $B$ surfaces at the  $m$-th 6DMA-BS  are mounted on a circular ring, where each surface can adjust its own (i) azimuth direction on the $x$--$y$ plane and (ii) tilt angle, as shown in Fig.~\ref{fig:circular_model}.

Accordingly, we parameterize the rotation vector of the $b$-th surface at the $m$-th 6DMA-BS  by a two-dimensional vector
\begin{equation}
	\mathbf{z}_{m,b}\triangleq[\varphi_{m,b},\,\theta_{m,b}]^{T}\in\mathbb{R}^{2\times 1},
\end{equation}
where $\varphi_{m,b}\in(0,2\pi]$ denotes the azimuth angle of the surface normal's projection on the $x$--$y$ plane, and $\theta_{m,b}$ denotes its tilt angle (subject to a physical range).
Stacking all surfaces, the rotation vector of the $m$-th  6DMA-BS can be written as
\begin{equation}
	\mathbf{z}_m \triangleq [\mathbf{z}_{m,1}^{T},\ldots,\mathbf{z}_{m,B}^{T}]^{T}\in\mathbb{R}^{2B\times 1},
\end{equation}
and the global rotation vector is $\mathbf{z}\triangleq[\mathbf{z}_1^{T},\ldots,\mathbf{z}_M^{T}]^{T}\in\mathbb{R}^{2BM \times 1}$.

The $m$-th 6DMA-BS serves $K_m$ users in its corresponding cell. Let $\mc{K}_m = \{k_{m,1}, \ldots, k_{m,K_m}\}$ denote the index set of the users served by the $m$-th 6DMA-BS. The index set of all users is denoted by $\mc{K} \triangleq \bigcup_{m=1}^{M} \mc{K}_m$, and the total number of users in all cells is denoted by $K = |\mc{K}| = \sum_{m=1}^{M} K_m$. The locations of the $K$ users are denoted by $\mathbf{X} \triangleq [\mathbf{x}_{1}, \dots, \mathbf{x}_{K}] \in \mbb{R}^{3\times K}$.

\begin{figure}[!t]
	\centering
	\includegraphics[width=3in]{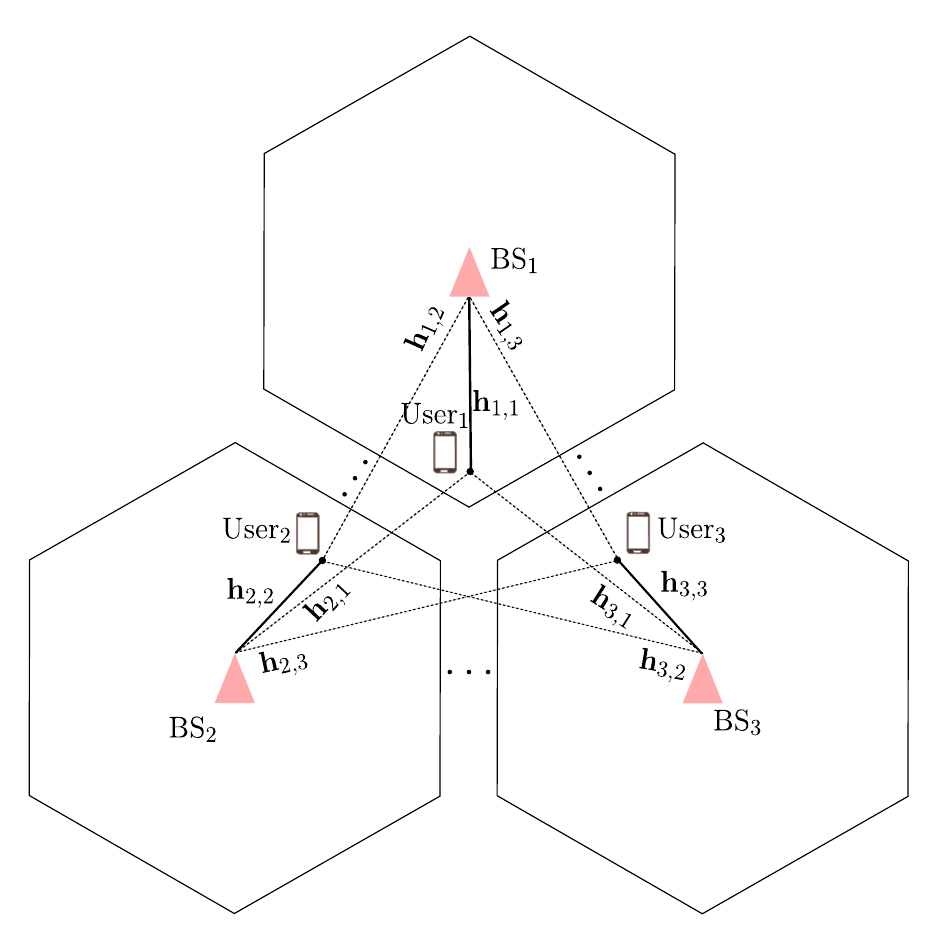}
\caption{Illustration of a multi-cell downlink network with $M=3$, where each 6DMA-BS serves its in-cell users and may cause ICI to users in other cells.  The gaps between the illustrated cells visually indicate their separation under frequency reuse.}
	\label{fig:channel_model}
\end{figure}

Let $\mathbf{f}_{m,k} \triangleq (\mathbf{x}_{k} - \mathbf{o}_m)/\|\mathbf{x}_{k} - \mathbf{o}_m\| \in \mathbb{R}^{3\times 1}$ denote the unit direction-of-departure (DoD) vector from the $m$-th 6DMA-BS to the user $k$ at position $\mathbf{x}_{k}$. As shown in Fig.~\ref{fig:channel_model}, the channel  from all the $B$ 6DMA surfaces of the $m$-th 6DMA-BS to user $k$ can be expressed as
\begin{align}
	\mbf{h}_{m,k}^H = [(\mbf{h}_{m,k}^1)^H,\cdots,(\mbf{h}_{m,k}^B)^H] \in \mbb{C}^{1\times NB},
\end{align}
where $(\mbf{h}_{m,k}^b)^H \in \mbb{C}^{1\times N}$ denotes the channel from the $b$-th 6DMA surface of the $m$-th 6DMA-BS to the user $k$. 
For the purpose of exposition, we consider a narrow-band system with flat-fading single-path channels in the far field, thus $(\mbf{h}_{m,k}^b)^H$ can be expressed as\footnote{The single-path channel model is adopted for simplicity. Under a general multipath channel, $(\mbf{h}_{m,k}^b)^H$ can be generalized as $\sum_{\ell=1}^{L_{m,k}}\sqrt{g(\mbf{z}_{m,b},\mbf{f}_{m,k,\ell})}\,\mathbf{a}^H(\mbf{z}_{m,b},\mbf{f}_{m,k,\ell})\,\xi_{m,k,\ell}$, where $L_{m,k}$ denotes the number of propagation paths from the $m$-th 6DMA-BS to user $k$, $\mbf{f}_{m,k,\ell}\in\mathbb{R}^{3\times 1}$ denotes the unit DoD vector of the $\ell$-th path, and $\xi_{m,k,\ell}$ denotes the corresponding complex path coefficient. }
\begin{align}
	(\mbf{h}_{m,k}^b)^H &=  \sqrt{g(\mbf{z}_{m,b},\mbf{f}_{m,k})}\mathbf{{a}}^H(\mbf{z}_{m,b},\mbf{f}_{m,k})  \xi_{m,k}, \label{hkb}
\end{align}
where  $\xi_{m,k}$, $m \in \mc{M}$, $k \in \mc{K}$, denotes the complex channel coefficient of the propagation path from the $m$-th 6DMA-BS to the $k$-th user;  $g(\mbf{z}_{m,b},\mbf{f}_{m,k})$ 
and $\mathbf{{a}}^H(\mbf{z}_{m,b},\mbf{f}_{m,k}) \in \mbb{C}^{1\times N}$, $b \in \mc{B}$, respectively denote the  antenna gain and  steering vector of the $b$-th 6DMA surface. In general, the antenna gain $g(\mbf{z}_{m,b},\mbf{f}_{m,k})$ and the steering vector $\mathbf{{a}}^H(\mbf{z}_{m,b},\mbf{f}_{m,k})$ both depend on the rotation vector $\mbf{z}_{m,b}$ of the $b$-th 6DMA surface and the DoD $\mbf{f}_{m,k}$. {Following the channel model  in \cite{6DMA_TWC}}, we have
\begin{align}
	g(\mbf{z}_{m,b},\mbf{f}_{m,k}) &= G\left(\mbf{R}^{-1}(\mbf{z}_{m,b}) \mbf{f}_{m,k}\right), \label{antenna_gain}\\
	\mathbf{{a}}^H(\mbf{z}_{m,b},\mbf{f}_{m,k}) &= [e^{j\frac{2\pi}{\lambda}\mbf{f}_{m,k}^{\top} \mbf{r}_{m,b,1}},\cdots,e^{j\frac{2\pi}{\lambda}\mbf{f}_{m,k}^{\top} \mbf{r}_{m,b,N}}] \label{steering_vec_def},
\end{align}
where $\mbf{R}(\mbf{z}_{m,b})$ is the rotation matrix in $Z$-$Y$ order corresponding to rotation vector $\mbf{z}_{m,b}$, 
which is given by \cite{rotation_matrix}
\begin{equation}
	\mathbf{R}\!\left(\mathbf{z}_{m,b}\right)
	=
	\begin{bmatrix}
		c_{\varphi_{m,b}}c_{\theta_{m,b}} & -s_{\varphi_{m,b}} & \;\;c_{\varphi_{m,b}}s_{\theta_{m,b}}\\
		s_{\varphi_{m,b}}c_{\theta_{m,b}} & \;\;c_{\varphi_{m,b}} & \;\;s_{\varphi_{m,b}}s_{\theta_{m,b}}\\
		-s_{\theta_{m,b}} & 0 & c_{\theta_{m,b}}
	\end{bmatrix},
\end{equation}
with $c_x \triangleq \cos(x)$ and $s_x \triangleq \sin(x)$ for notational brevity.
In \eqref{antenna_gain}, it can be shown that  $\mbf{R}^{-1}(\mbf{z}_{m,b}) \mbf{f}_{m,k} = \mbf{R}^\top(\mbf{z}_{m,b}) \mbf{f}_{m,k}$, which denotes the DoD $\mbf{f}_{m,k}$ after being projected to the  local Cartesian coordinate system (CCS) of the $b$-th 6DMA surface with rotation $\mbf{z}_{m,b}$; the function $G(\cdot)$ denotes the effective gain of antennas of each 6DMA surface in terms of the DoD in its local CCS (to be specified in Section V  based on the practical antenna radiation pattern adopted). In \eqref{steering_vec_def}, $\lambda$ denotes the wavelength of the carrier wave, and $\mbf{r}_{m,b,n} \in\mbb{R}^{3\times 1}$, $n\in\mc{N}$, represents the location of the $n$-th antenna on the $b$-th 6DMA surface with the rotation vector $\mbf{z}_{m,b}$ in the CCS of the $m$-th 6DMA-BS, i.e., 
\begin{align}
	\mbf{r}_{m,b,n} = \mathbf{q}_{m,b}\!\left(\mathbf{z}_{m,b}\right) + \mbf{R}(\mbf{z}_{m,b}) \bar{\mbf{r}}_n, \label{comformal_trans}
\end{align}
where the position of the $b$-th 6DMA surface on the circular track of the $m$-th 6DMA-BS, $\mbf{q}_{m,b} \in\mbb{R}^{3\times 1}$, is given by
\begin{equation}
	\mathbf{q}_{m,b}\!\left(\mathbf{z}_{m,b}\right)
	=
	\begin{bmatrix}
		\rho_m\cos\varphi_{m,b}\\
		\rho_m\sin\varphi_{m,b}\\
		0
	\end{bmatrix},
\end{equation}
where $\rho_m$ denotes the radius of the circular track of the $m$-th 6DMA-BS, and the $n$-th antenna of a 6DMA surface in its local CCS, $\bar{\mathbf r}_n\in\mathbb R^{3\times 1}$, is predefined  based on the  practical geometry of 6DMA surfaces (e.g., uniform planar array (UPA)).

 \subsection{Signal Model}

Let $\mathbf{w}_k \in \mathbb{C}^{NB\times 1}$ denote the transmit precoding vector for user $k$ at its serving 6DMA-BS $m(k)$, where $m(k)$ denotes the index of the serving cell of the $k$-th user. Define
$\mathbf{W}_m \triangleq [\mathbf{w}_{k_{m,1}},\ldots,\mathbf{w}_{k_{m,K_m}}]\in\mathbb{C}^{NB\times K_m}$ as the precoding matrix at the $m$-th 6DMA-BS.
Accordingly, the transmit signal of the $m$-th 6DMA-BS is given by
\begin{equation}
	\mathbf{s}_m=\sum_{k\in\mathcal{K}_m}\mathbf{w}_k s_k \in\mathbb C^{NB \times 1},
\end{equation}
where $s_k$ is the data symbol intended for user $k$ with $\mathbb{E}[|s_k|^2]=1$.
The precoders are subject to the per-BS transmit power constraint
\begin{equation}
	\sum_{k\in\mathcal{K}_m}\|\mathbf{w}_k\|_2^2 \le P_{\max},\quad \forall m\in \mc{M},
\end{equation}
where  $P_{\max}$ denotes the  transmit power budget of each 6DMA-BS.
Thus, the received baseband signal at user $k$ is given by
\begin{align}
	y_k
	&= \mbf{h}_{m(k),k}^{H}\mbf{w}_k s_k
	+ \sum_{\substack{k'\in \mc K_{m(k)}\\ k'\neq k}} \mbf{h}_{m(k),k}^{H}\mbf{w}_{k'} s_{k'}
	+ \nonumber\\  &\sum_{\substack{m'\neq m(k)}} \sum_{k'\in \mc K_{m'}} \mbf{h}_{m',k}^{H}\mbf{w}_{k'} s_{k'}
	+ n_k,
\end{align}
where $n_k\sim\mathcal{CN}(0,\sigma^2)$ denotes the additive Gaussian noise. Consequently, the instantaneous signal to interference-plus-noise ratio (SINR) of user $k$ is given by
\begin{align}
	&\gamma_k = \nonumber \\
	&\frac{ 
		\left| \mathbf{h}_{m(k),k}^H \mathbf{w}_k \right|^2 
	}{
		\sum\limits_{\substack{k' \in \mathcal{K}_{m(k)}\\ k' \neq k}} 
		\left| \mathbf{h}_{m(k),k}^H \mathbf{w}_{k'} \right|^2
		+ \sum\limits_{m' \neq m(k)} \sum\limits_{k' \in \mathcal{K}_{m'}} 
		\left| \mathbf{h}_{m',k}^H \mathbf{w}_{k'} \right|^2
		+ \sigma^2	}.
	\label{eq:SINR_inst}
\end{align}
Then, the corresponding achievable rate of user $k$ in bits/second/Hertz is $\log_2 \left( 1 + {\gamma}_k \right)$.
\subsection{Problem Formulation}
In this paper, we aim to maximize the average weighted
sum-rate of all the users by jointly optimizing the short-term
transmit precoding at the BSs and long-term antenna orientations,
subject to the maximum transmit power constraint at the BSs.
Accordingly, our considered two-timescale optimization problem is formulated as follows:
 \begin{subequations}
 	\label{P1}
 	\begin{align}
 	\text{(P1):} \quad&\max_{\mathbf{z}\in\mathcal{Z}} ~ \mathbb{E} \left\{ 
 		\max_{\{\mathbf{W}_m\}_{m=1}^M} \sum_{k \in\mc{K}} 
 		\alpha_k \log_2 \left( 1 + {\gamma}_k \right)  \right\} \label{eq:p1_opt_obj} \\
 		&\text{s.t.} ~~ \sum_{k\in\mc{K}_m}\|\mathbf{w}_k\|_2^2 \leq P_{\max}, \quad \forall m, \label{eq:p1_power_constraint}
 	\end{align}
 \end{subequations}
 where $\alpha_k > 0$ is the predefined user-rate weight, and the expectation is taken over the user distribution and small-scale channel fading. 
 To avoid overlap among different 6DMA surfaces on the circular track, we enforce a minimum angular separation between the azimuth directions of any two surfaces.
 Specifically, the feasible set of rotations is defined as
 \begin{equation}
 	\mathcal{Z} \triangleq 
 	\Big\{\mathbf{z}=[\mathbf{z}_1^T,\ldots,\mathbf{z}_M^T]^T \ \Big|\ 
 	\mathbf{z}_m\in\mathcal{Z}_m,\ \forall m\in\mathcal{M}\Big\},
 	\label{eq:Z_global}
 \end{equation}
 where
 \begin{equation}
 	\mathcal{Z}_m \triangleq 
 	\left\{\mathbf{z}_m ~\middle|~
 	\begin{aligned}
 		& \varphi_{m,b}\in(0,2\pi],\ \theta_{m,b}\in[\theta_{\min},\theta_{\max}],\ \forall b,\\
 		& d\!\left(\varphi_{m,b},\varphi_{m,b'}\right)\ge \Delta\phi,\ \forall\,1\le b<b'\le B
 	\end{aligned}
 	\right\},
 	\label{eq:Z_local}
 \end{equation}
with $d(\varphi,\psi)\triangleq \min\{|\varphi-\psi|,\,2\pi-|\varphi-\psi|\}$ denoting the circular angular distance, $\Delta\phi>0$ denoting the prescribed minimum azimuth separation (guard angle) between the $x$--$y$ plane projection directions of any two surfaces to avoid overlap, and $\theta_{\min}$ and $\theta_{\max}$ denoting the minimum and maximum allowable tilt angles of each 6DMA surface, respectively.

\section{Centralized Cooperation: Offline Optimization}

We first consider the  centralized cooperation framework, where a network controller coordinates all the $M$ cells with access to the global CSI of the entire network. In particular, the controller jointly optimizes the long-term 6DMA rotation vector $\mathbf z \in \mc Z$ and the short-term downlink precoders $\{\mathbf W_m\}_{m=1}^M$ to maximize the network-wide average weighted sum-rate in \eqref{P1}. 
This centralized formulation preserves the cellular structure, where each 6DMA-BS serves its in-cell users, while enabling full inter-cell coordination through global information sharing.
It is worth noting that the expectation in \eqref{P1} is generally intractable to evaluate in closed form. To obtain an offline solution, we adopt a sample-average approximation (SAA) approach and develop an alternating optimization (AO) procedure.

\subsection{Offline Solution via SAA and Particle Swarm Optimization (PSO)-based Long-Term Search}

We generate $S$ independent realizations of the random network state, including user locations and small-scale channel fading, denoted by $\{\omega^{(s)}\}_{s=1}^S$, where each sample $\omega^{(s)}$ induces a corresponding set of channel vectors $\{\mathbf h_{m,k}^{(s)}\}$ according to the channel model in Section~II.
For any given $\mathbf z\in\mc Z$, define the instantaneous weighted sum-rate maximization value for sample $s$ as
\begin{subequations}
	\begin{align}
		J^{(s)}(\mathbf z)
		\triangleq
		\max_{\{\mathbf W_m\}_{m=1}^M} \quad
		& \sum_{k\in\mc K}\alpha_k \log_2\!\left(1+\gamma_k^{(s)}\right) \\
		\text{s.t.} \quad
		& \sum_{k\in\mc K_m}\|\mathbf w_k\|_2^2\le P_{\max},~\forall m,
	\end{align}\label{eq:g_sample}
\end{subequations}
\hspace{-0.9em} where $\gamma_k^{(s)}$ is obtained from \eqref{eq:SINR_inst} by replacing $\{\mathbf h_{m,k}\}$ with $\{\mathbf h_{m,k}^{(s)}\}$.
For each fixed $(\mathbf z,s)$, the inner problem \eqref{eq:g_sample} is a standard weighted sum-rate maximization under per-BS power constraints and can be efficiently handled via the celebrated weighted minimum mean-square error (WMMSE) reformulation \cite{Christensen2008WMMSE,Shi2011WMMSE}.
In particular, \eqref{eq:g_sample} is equivalently transformed into a WMMSE minimization by introducing auxiliary user-side scalar equalizers and mean-square error (MSE) weights. The resulting alternating procedure monotonically improves the objective value and converges to a stationary point, yielding an approximate optimizer $\{\mathbf W_m^{(s)}(\mathbf z)\}$ and the corresponding value $J^{(s)}(\mathbf z)$.

Then, the SAA objective of \eqref{P1} is given by
\begin{equation}
	\widehat{J}_S(\mathbf z)
	\triangleq
	\frac{1}{S}\sum_{s=1}^S J^{(s)}(\mathbf z),
	\label{eq:SAA_obj}
\end{equation}
and the centralized offline solution can be approximated by
\begin{equation}
	\max_{\mathbf z\in\mc Z}\ \widehat{J}_S(\mathbf z).
	\label{eq:SAA_problem}
\end{equation}

Problem \eqref{eq:SAA_problem} preserves the two-timescale structure of \eqref{P1}. For each long-term variable $\mathbf z$, the inner short-term optimization \eqref{eq:g_sample} is first solved for all $s=1,\ldots,S$, and the long-term variable is then updated according to the sample-average objective \eqref{eq:SAA_obj}. To keep the long-term solver consistent with the decentralized benchmark developed later, we adopt particle swarm optimization as a practical solver for \eqref{eq:SAA_problem} \cite{KennedyEberhart1995PSO}. For subsequent complexity analysis, let $N_{\mathrm{sw}}^{\mathrm c}$ and $T_{\mathrm{PSO}}^{\mathrm c}$ denote the swarm size and the maximum number of PSO iterations for the centralized offline search, respectively.

Specifically, we equivalently minimize the negative sample-average objective
\begin{equation}
	F_S(\mathbf z)
	=
	-\widehat{J}_S(\mathbf z),
	\label{eq:SAA_obj_pso}
\end{equation}
and incorporate the constraint $\mathbf z\in\mc Z$ through a penalty-based formulation
\begin{equation}
	\tilde{F}_S(\mathbf z)
	=
	-\widehat{J}_S(\mathbf z)
	+
	\kappa_{\mathrm c}\Psi_{\mathrm c}(\mathbf z),
	\label{eq:SAA_penalty}
\end{equation}
where $\Psi_{\mathrm c}(\mathbf z)$ is a nonnegative penalty function characterizing the violation of the feasible set $\mc Z$, and $\kappa_{\mathrm c}>0$ is the associated penalty coefficient. By construction, $\Psi_{\mathrm c}(\mathbf z)=0$ holds for any feasible $\mathbf z\in\mc Z$, while $\Psi_{\mathrm c}(\mathbf z)>0$ otherwise.

Starting from a feasible initialization $\mathbf z^{(0)}\in\mc Z$, one particle is warm-started from $\mathbf z^{(0)}$, while the remaining particles are randomly initialized. A standard PSO routine is then applied to iteratively update the particle positions and velocities, where the objective value of each particle is evaluated through \eqref{eq:SAA_penalty}. For each candidate long-term vector, this evaluation requires solving \eqref{eq:g_sample} for all $S$ channel samples and then forming the sample-average objective in \eqref{eq:SAA_obj}. After the stopping criterion is met, the best feasible particle returned by the swarm is adopted as the approximate centralized offline solution to \eqref{eq:SAA_problem}.

\subsection{Complexity and Practical Limitations}
We analyze the computational complexity of solving \eqref{eq:SAA_problem} via the offline PSO-based long-term search in Section~III-A. Let $N_{\mathrm{sw}}^{\mathrm c}$ and $T_{\mathrm{PSO}}^{\mathrm c}$ denote the swarm size and the maximum number of PSO iterations for the centralized offline search, respectively. Let $T_{\mathrm{WMMSE}}$ denote the number of WMMSE iterations used to obtain a stationary solution of the short-term problem \eqref{eq:g_sample} for each sample $s$.

\textit{Short-term optimization complexity:}
For fixed $(\mathbf z,s)$, the WMMSE-based solver for \eqref{eq:g_sample} requires, in each iteration, solving linear system equations whose dimensions scale with the number of transmit dimensions per BS \cite{Shi2011WMMSE,GolubVanLoan2013}. This leads to a per-iteration arithmetic complexity that scales as $\mc O((NB)^3)$ per BS. Summing over all $M$ BSs yields the per-iteration complexity $\mc O\!\left(M(NB)^3\right)$. Therefore, solving \eqref{eq:g_sample} for one sample requires
\[
\mc O\!\left(T_{\mathrm{WMMSE}}\,M(NB)^3\right).
\]

\textit{SAA evaluation and long-term update:}
Evaluating $\widehat J_S(\mathbf z)$ in \eqref{eq:SAA_obj} requires solving \eqref{eq:g_sample} for all $S$ samples, resulting in
\[
\mc O\!\left(S\,T_{\mathrm{WMMSE}}\,M(NB)^3\right).
\]
In the centralized offline design, the long-term variable $\mathbf z$ is optimized by PSO. For each particle and each PSO iteration, one objective evaluation is required. Since the penalty term in \eqref{eq:SAA_penalty} is of lower complexity than the SAA objective evaluation, the per-particle cost is dominated by computing $\widehat J_S(\mathbf z)$. Define the effective long-term search count of the centralized PSO solver as
$
	{T_{\mathrm{PSO}}^{\mathrm c}}^\prime
	\triangleq
	N_{\mathrm{sw}}^{\mathrm c} T_{\mathrm{PSO}}^{\mathrm c}.
	\label{eq:Tpsoc_eff}
$
Consequently, the total complexity of the offline procedure scales as
\begin{equation}
	\mc O\!\left(
	{T_{\mathrm{PSO}}^{\mathrm c}}^\prime
	S\,
	T_{\mathrm{WMMSE}}\,
	M(NB)^3
	\right).
	\label{eq:total_cost_simplified}
\end{equation}

The key implication of \eqref{eq:total_cost_simplified} is its explicit dependence on the number of cells, $M$. In particular, the short-term WMMSE updates must be carried out for all $M$ BSs for each sample and each long-term iteration, which leads to a total cost that grows with $M$ and can quickly become prohibitive as the network scales. Beyond high computational complexity, the centralized cooperation framework imposes stringent system requirements. In particular, acquiring and maintaining global CSI and coordinating the joint design across all cells necessitate substantial information exchange over the backhaul and a powerful network controller, which are difficult to implement   in large-scale networks.

The above complexity and coordination overhead motivate the development of more scalable cooperation architectures with reduced backhaul signaling and improved computational efficiency, which will be pursued in Section~IV. Meanwhile, the centralized offline solution developed in this section will be used as a performance upper bound in the simulation study later to analyze the performance gap of the proposed decentralized cooperation design from it.

\section{Decentralized Cooperation: Distributed Two-Timescale Algorithm}
In this section, we present our proposed distributed two-timescale algorithm for solving problem (P1). We adopt a frame-based protocol where the short-term precoders are updated per transmission block using instantaneous CSI, while the IPC thresholds (to be explicitly defined) and long-term rotation vectors are optimized over the transmission frame based on locally estimated statistical CSI at individual BSs and limited inter-BS coordination.

In the decentralized coordination system, the $m$-th 6DMA-BS solves the following per-BS problem
\begin{subequations}
	\begin{align}
		\text{(P2):}\quad \max_{\mathbf{z}_m \in \mc{Z}_m}\quad 
		& \mathbb{E}\!\left\{
		\max_{\mathbf W_m}
		\sum_{k\in\mc K_m}\alpha_k \log_2\!\left(1+\gamma_k\right)
		\right\} \label{eq:p2_opt_obj} \\
		\text{s.t.}\quad 
		& \sum_{k\in\mc{K}_m}\|\mathbf{w}_k\|_2^2 \le P_{\max}, \label{eq:p2_power_constraint}
	\end{align}
\end{subequations}
where the expectation is taken with respect to the local channel statistics associated with 6DMA-BS $m$ only.

It is worth noting that problem (P1) is a network-wide two-timescale optimization. The rotation vector $\mathbf z_m$ and the precoders $\mathbf W_m$ of 6DMA-BS $m$ affect not only the rates of its in-cell users but also the rates of out-of-cell users through ICI.
In contrast, (P2) is a local design that optimizes the in-cell weighted sum-rate under the per-BS transmit power constraint.
However, the SINR terms in (P2) are still coupled through ICI. As such, evaluating $\{\gamma_k\}_{k\in\mc K_m}$ for short-term precoder updates generally requires information about the instantaneous interference created by other BSs, which may entail substantial CSI feedback from in-cell users to their serving BS as well as CSI  exchange among BSs.
Moreover, the long-term expectation in (P2) mainly depends on interference statistics contributed by neighboring BSs. Without additional coordination, each BS would need access to cross-cell statistical information to accurately update $\mathbf z_m$, which undermines the goal of limiting inter-BS signaling.
Therefore, independently solving (P2) at all 6DMA-BSs is, in general, neither equivalent to (P1) nor sufficient to remove coordination overhead, and may lead to a non-negligible performance gap with respect to the network optimum, especially in interference-limited regimes.

\begin{figure*}[t]
	\centering
	\includegraphics[width=6in]{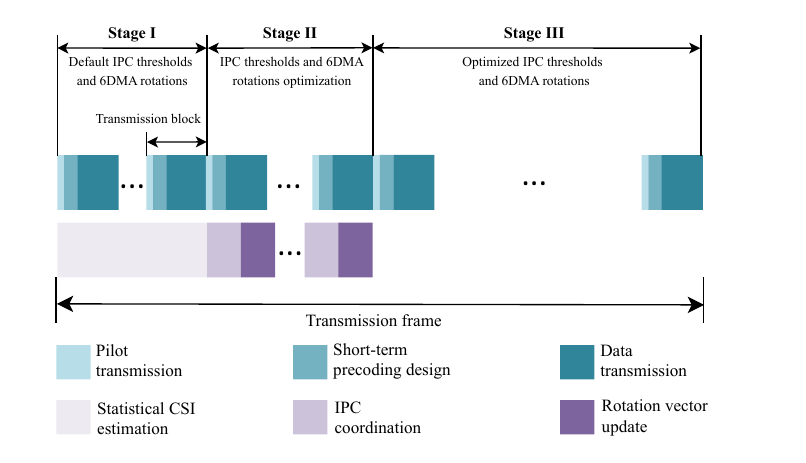} 
\caption{Illustration of the proposed two-timescale transmission protocol within one transmission frame. In each transmission block, pilot transmission is followed by short-term precoding design and data transmission based on the currently adopted IPC thresholds and 6DMA rotation vectors. Over the transmission frame, statistical CSI is accumulated across the transmission blocks in Stage I, while the communication is carried out with the default IPC thresholds and 6DMA rotations. Stage II corresponds to communication with IPC thresholds and 6DMA rotations that are iteratively updated during the optimization process based on the statistical CSI acquired in Stage I. After sufficient updates, Stage III corresponds to communication with the optimized IPC thresholds and 6DMA rotations.}
	\label{fig:protocol}
\end{figure*}

To address the above limitations while retaining a distributed implementation, we further introduce an  IPC-assisted design in this section.
Specifically, we augment the local problem (P2) with additional constraints that upper-bound the interference leakage from 6DMA-BS $m$ to a selected set of neighboring-cell users, resulting in the IPC-constrained short-term formulation in (P4).
These IPC constraints alleviate the ICI coupling in a controlled manner by replacing the need for acquiring detailed cross-cell CSI/statistical information with specifying a small number of IPC thresholds.
The thresholds are coordinated among neighboring 6DMA-BSs via limited signaling, such as exchanging candidate IPC threshold values and the corresponding scalar local utility values, and each BS then performs its local short-term precoder update and long-term rotation update subject to the agreed IPC levels.
As a result, the IPC-constrained local optimization  reduces coordination overhead and also better mitigates the impact of each BS's local design on neighboring cells, yielding a distributed solution that empirically approaches the network-wide performance of (P1).
Fig.~\ref{fig:protocol} illustrates the proposed frame-based protocol.

\subsection{Short-Term Optimization}

In the short-term timescale, the rotation vector of each 6DMA-BS is  {fixed}.
Specifically, the $m$-th 6DMA-BS takes the current rotation vector $\mathbf{z}_m$ as given, where $\mathbf{z}_m\in\mathcal{Z}_m$ is ensured by the long-term rotation update.
With fixed $\mathbf{z}_m$, the $m$-th 6DMA-BS  optimizes only the downlink precoding vectors $\mbf{W}_m$ by solving
\begin{subequations}
	\begin{align}
	\text{(P3):}\quad	\max_{\mbf{W}_m} \quad
		& \sum_{k\in\mathcal{K}_m} \alpha_k \log_2\!\left(1+\gamma_k\right) \label{eq:p3_opt_obj} \\
		\text{s.t.} \quad
		& \sum_{k\in\mathcal{K}_m}\|\mathbf{w}_k\|_2^2 \le P_{\max}. \label{eq:p3_power_constraint}
	\end{align}
	\label{P_short_term_ori}
\end{subequations}

\begin{figure}[t]
 	\centering
 	\includegraphics[width=3in]{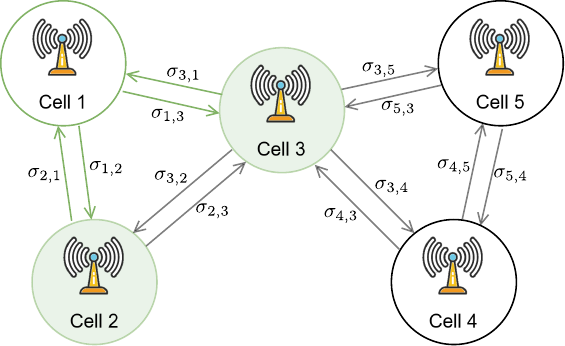} 
 		\caption{Graph-based illustration of adjacent-cell sets and IPC. Each node represents a 6DMA-BS cell, and each directed edge labeled by $\sigma_{m,n}$ denotes the IPC threshold that limits the interference leakage from cell $m$ to its adjacent cell $n$. The highlighted example shows that the adjacent-cell set of Cell 1 is $\mathcal{A}_1=\{2,3\}$, and hence Cell 1 imposes IPCs $\sigma_{1,2}$ and $\sigma_{1,3}$ toward Cells 2 and 3, respectively, and vice versa.}
 	\label{BS_top}
 	\end{figure}

However, as aforementioned, due to the lack of CSI of other cells, the term
$\sum\limits_{m' \neq m(k)} \sum\limits_{k' \in \mathcal{K}_{m'}} 
\left| \mathbf{h}_{m',k}^H \mathbf{w}_{k'} \right|^2$
in \eqref{eq:SINR_inst} is generally unavailable to the $m$-th 6DMA-BS under a decentralized implementation.
Accurately evaluating this ICI term would require substantial inter-BS coordination (e.g., exchanging cross-cell CSI and/or precoding-related information) over the backhaul, which is undesirable in large-scale networks.
As a result, although the problem remains solvable in principle, a direct implementation based on exact interference knowledge is impractical without high coordination overhead.
 To tackle this issue, we introduce an IPC to mitigate the power leakage of cell $m\in \mc{M}$ to its adjacent cells. Let $\mc{A}_m$ denote the set of adjacent co-channel cells coordinated with cell $m$, as shown in Fig.~\ref{BS_top}.
Then, we impose\footnote{We assume that the channels from each BS to the users in its adjacent cells can be estimated via uplink training. }
 \begin{align}
 	\sum_{k \in \mc{K}_m} |\mathbf{h}_{m,k'}^H \mbf{w}_k|^2 \le \sigma_{m,n},~\forall k' \in \mc{K}_{n},
 \end{align}
 where  $\sigma_{m,n}$ denotes the IPC threshold from cell $m$ to $n$, which is to be determined by the coordination scheme proposed in {Section~IV}. Thus, we can lower-bound the SINR $\gamma_k$  by 
 \begin{align}
	&\hat{\gamma}_k = \nonumber \\
&\frac{ 
	\left| \mathbf{h}_{m,k}^H \mathbf{w}_k \right|^2 
}{
	\sum\limits_{\substack{k' \in \mathcal{K}_{m}\\ k' \neq k}} 
	\left| \mathbf{h}_{m,k}^H \mathbf{w}_{k'} \right|^2
	+ \sum\limits_{n \in \mc{A}_{m}}  \sigma_{n,m}
	+ \sigma^2	},~ k\in\mc{K}_m. \label{SINR_bound}
 \end{align}
 	
 By replacing $\gamma_k$ by $\hat{\gamma}_k$, the approximate achievable rate of user $k$,  $\log_2 (1 + \hat{\gamma}_k)$, becomes independent of the CSI acquisition of other cells, and the $m$-th BS can alternatively solve the following problem in a  distributed manner:
     \begin{subequations}
 	\begin{align}
 	\text{(P4):}\quad	\max_{\mbf{W}_m} &\sum_{k\in \mc{K}_m} 
 		\alpha_k \log_2 \left( 1 + \hat{\gamma}_k \right) \label{eq:p4_opt_obj} \\
 		\text{s.t.} ~~~~&  \sum_{k\in \mc{K}_m} |\mathbf{h}_{m,k'}^H \mbf{w}_k|^2  \le \sigma_{m,n},~\forall k' \in \mc{K}_n,~\forall n \in \mc{A}_m, \label{IPC_constraint}\\
 		~~~~&\sum_{k\in\mc{K}_m} \|\mbf{w}_k\|_2^2 \leq P_{\max}. \label{total_power_constraint}
 	\end{align}
 	\label{P_short_term_new}
 \end{subequations}
 
 Following the WMMSE equivalence \cite{Shi2011WMMSE}, maximizing the weighted sum-rate is equivalent to minimizing a weighted sum of MSEs. Specifically, with scalar receive filters $\{u_k\}$ and positive weights $\{v_k\}$, the MSE of estimating $s_k$ from the received signal $y_k$ using $u_k$ is given by
 \begin{align}
e_k &= \mathbb{E}\{|u_k^* y_k - s_k|^2\} \nonumber\\
&= 1 - 2\Re\{u_k^*\mathbf{h}_{m,k}^H\mathbf{w}_k\}
+ |u_k|^2\!\left(\sum_{j\in\mathcal{K}_m}|\mathbf{h}_{m,k}^H\mathbf{w}_j|^2 + c_k\right),
 	\label{eq:mse_def}
 \end{align}
where $c_k$, $\forall k \in \mc{K}_m$, collects the noise and the upper-bounded ICI terms of the $k$-th user, given by
 \begin{align}
 	c_k = \sum\limits_{n \in \mc{A}_{m}}  \sigma_{n,m}
 	+ \sigma^2,~\forall k \in \mc{K}_m.
 \end{align}
For fixed $\mathbf{W}_m$, the minimum mean-square error (MMSE) receiver and the corresponding optimal MSE weight are
\begin{align}
	u_k &= 
	\frac{\mathbf{h}_{m,k}^H \mathbf{w}_k}
	{\sum_{j\in\mathcal{K}_m}|\mathbf{h}_{m,k}^H\mathbf{w}_j|^2 + c_k},
	\qquad
	v_k = \frac{1}{e_k}.
	\label{eq:uv_update}
\end{align}
 Then, for fixed $\{u_k,v_k\}$, the BS minimizes $\sum_{k\in\mathcal K_m} \alpha_k(v_k e_k - \log v_k)$, which reduces to a convex quadratically-constrained quadratic program (QCQP).
 
For notational convenience, we rewrite the IPC constraints in \eqref{IPC_constraint} in a unified form indexed by
$q=1,\ldots,Q_m$, where $Q_m \triangleq \sum_{n\in\mathcal{A}_m}|\mathcal{K}_n|$.
Specifically, we establish a one-to-one mapping between the index $q$ and an adjacent-cell user pair $(n(q),k'(q))$
with $n(q)\in\mathcal{A}_m$ and $k'(q)\in\mathcal{K}_{n(q)}$, and define
\begin{equation}
	\boldsymbol{\ell}_{m,q} \triangleq \mathbf{h}_{m,k'(q)}, 
	\qquad
	\bar{\sigma}_{m,q} \triangleq \sigma_{m,n(q)} .
	\label{eq:g_s_mapping}
\end{equation}
Thus the IPC constraint \eqref{IPC_constraint} can be compactly written as
\begin{equation}
	\sum_{k\in\mathcal{K}_m}\big|\boldsymbol{\ell}_{m,q}^H\mathbf{w}_k\big|^2 \le \bar{\sigma}_{m,q},
	\quad q=1,\ldots,Q_m.
	\label{eq:ipc_compact_mq}
\end{equation}

Next, we introduce dual variables $\{\lambda_{m,q}\}_{q=1}^{Q_m}$ with $\lambda_{m,q}\ge 0$ for the IPC constraints
\eqref{eq:ipc_compact_mq} and $\mu_m\ge 0$ for the total power constraint \eqref{total_power_constraint}. 
For convenience, define
\begin{equation}
	\chi_k \triangleq \alpha_k v_k |u_k|^2,
	\qquad
	\mathbf{b}_{m,k} \triangleq \alpha_k v_k u_k \mathbf{h}_{m,k},
	\quad \forall k\in\mathcal{K}_m,
	\label{eq:ak_bk_m}
\end{equation}
and the Hermitian matrix
\begin{equation}
	\mathbf{C}_m(\boldsymbol{\lambda}_m)
	\triangleq
	\sum_{i\in\mathcal{K}_m} \chi_i  \mathbf{h}_{m,i}\mathbf{h}_{m,i}^H
	+
	\sum_{q=1}^{Q_m} \lambda_{m,q}\, \boldsymbol{\ell}_{m,q}\boldsymbol{\ell}_{m,q}^H,
	\label{eq:C_lambda_m}
\end{equation}
where $\boldsymbol{\lambda}_m \triangleq [\lambda_{m,1},\ldots,\lambda_{m,Q_m}]^T$ collects all IPC dual variables of cell $m$.
Then, the Karush–Kuhn–Tucker (KKT) conditions of the transmit-update subproblem yield the following closed-form beamformer structure:
\begin{equation}
	\mathbf{w}_k(\boldsymbol{\lambda}_m,\mu_m)
	=
	\left(\mathbf{C}_m(\boldsymbol{\lambda}_m)+\mu_m\mathbf{I}\right)^{-1}\mathbf{b}_{m,k},
	\quad \forall k\in\mathcal{K}_m.
	\label{eq:w_closed_form_m}
\end{equation}
For fixed $\boldsymbol{\lambda}_m$, the mapping
\(
\mu_m \mapsto \sum_{k\in\mathcal{K}_m}\|\mathbf{w}_k(\boldsymbol{\lambda}_m,\mu_m)\|_2^2
\)
is monotonically decreasing. Hence, $\mu_m$ can be efficiently found by bisection \cite{Boyd2004Convex} (with $\mu_m=0$ if the power constraint is inactive).

To enforce IPC feasibility, we update $\boldsymbol{\lambda}_m$ by projected subgradient ascent on the dual function:
\begin{align}
	\lambda_{m,q}^{(\iota+1)}
	&=
	\Big[\lambda_{m,q}^{(\iota)}+\delta_\iota\big(I_{m,q}(\mathbf{W}_m)-\bar{\sigma}_{m,q}\big)\Big]_+,
	\label{eq:lambda_update_m}\\
	I_{m,q}(\mathbf{W}_m)
	&\triangleq
	\sum_{k\in\mathcal{K}_m}\big|\boldsymbol{\ell}_{m,q}^H\mathbf{w}_k\big|^2,
	\label{eq:Imq_def}
\end{align}
where  $\iota$ denotes the inner dual-iteration index, $[\cdot]_+ \triangleq \max(\cdot,0)$ is applied elementwise, and $\delta_{\iota}$ is a diminishing step size.
In practice, only a small number of inner dual updates for $(\boldsymbol{\lambda}_m,\mu_m)$ is required per WMMSE iteration, resulting in a computationally efficient solver without generic QCQP toolboxes.
 
\subsection{Long-Term Optimization}
In this subsection, we propose a decentralized coordination scheme to update the IPC thresholds and the rotations of the $M$ 6DMA-BSs.

First, let $\mathbf{H}_m$ collect the channels from the $m$-th 6DMA-BS to (i) its served users and (ii) the users in its adjacent cells, and $\boldsymbol{\sigma}_m$ collect all IPC thresholds associated with cell $m$.
Specifically, for each 6DMA-BS $m$, let the user indices in $\mathcal{K}_m$ and 
$\cup_{n\in\mathcal{A}_m}\mathcal{K}_n$, as well as the adjacent 6DMA-BS indices in 
$\mathcal{A}_m$, be arranged in a given order. Then, we define
\begin{align}
	\mathbf{H}_m^i &\triangleq \big[\mathbf{h}_{m,k}\big]_{k\in\mathcal{K}_m}\in \mathbb{C}^{NB\times K_m},\\
	\mathbf{H}_m^e &\triangleq \big[\mathbf{h}_{m,k}\big]_{k\in\cup_{n\in\mathcal{A}_m}\mathcal{K}_n}\in \mathbb{C}^{NB\times \sum_{n\in\mathcal{A}_m} K_n},\\
	\mathbf{H}_m   &\triangleq \big[\mathbf{H}_m^i,\mathbf{H}_m^e\big]\in \mathbb{C}^{NB\times \left(K_m+\sum_{n\in\mathcal{A}_m} K_n\right)} ,
\end{align}
and
\begin{equation}
	\boldsymbol{\sigma}_m
	\triangleq
	\left[
	\{\sigma_{m,n}\}_{n\in\mathcal{A}_m},
	\{\sigma_{n,m}\}_{n\in\mathcal{A}_m}
	\right]^T \in \mathbb{R}_+^{2|\mathcal{A}_m|\times 1}.
\end{equation}

Let $R_m(\mathbf{H}_m,\boldsymbol{\sigma}_m)$ denote the surrogate weighted sum-rate achieved by cell $m$ after solving \eqref{P_short_term_new}. 
We define the corresponding long-term (average) performance as
\begin{equation}
	\bar{R}_m(\mathbf{z}_m,\boldsymbol{\sigma}_m)\triangleq
	\mathbb{E}_{\mathbf{H}_m}\!\left\{R_m(\mathbf{H}_m,\boldsymbol{\sigma}_m)\right\},
\end{equation}
where the expectation is taken over the channel distribution of $\mathbf{H}_m$, which is parameterized by the 6DMA surface rotation $\mathbf{z}_m$.
The long-term rate $\bar{R}_m$ can be efficiently approximated via the SAA method.
Since each BS knows the channel distribution of its served users, it can generate a fixed set of $S$ independent and identically distributed (i.i.d.) realizations of $\mathbf{H}_m^i$, denoted by $\{\mathbf{H}_m^{i,(s)}\}_{s=1}^{S}$.
For the channels from the $m$-th 6DMA-BS to adjacent-cell users, $\mathbf{H}_m^e$, the required statistical channel information can be obtained during uplink transmission (e.g., \cite{6DMA_Qijun}), based on which we generate $\{\mathbf{H}_m^{e,(s)}\}_{s=1}^{S}$.
Let $\mathbf{H}_m^{(s)}\triangleq[\mathbf{H}_m^{i,(s)},\mathbf{H}_m^{e,(s)}]$.
Then,
\begin{align}
	\bar{R}_m(\mathbf{z}_m,\boldsymbol{\sigma}_m)&\approx
	\frac{1}{S}\sum_{s=1}^{S} R_m(\mathbf{H}_m^{(s)},\boldsymbol{\sigma}_m) \triangleq \widehat{R}_m(\mathbf{z}_m,\boldsymbol{\sigma}_m),
\end{align}
where $S$ is the number of Monte Carlo samples.
Importantly, the same set of samples is reused for all evaluations of $\bar{R}_m$, which improves numerical stability and ensures reproducibility.

The proposed distributed long-term optimization is implemented in an alternating-optimization (AO) manner. Let $t$ denote the AO outer-iteration index and  $T_{\max}$ denote the maximum number of AO iterations. At each AO iteration, the IPC thresholds and the BS rotation variables are updated alternately. Specifically, with the current rotation variable $\mathbf{z}^{(t)}
=
\big[(\mathbf{z}_1^{(t)})^T,\ldots,(\mathbf{z}_M^{(t)})^T\big]^T$   fixed, all BSs first perform the IPC coordination procedure to update the IPC thresholds, which yields the new threshold set $\{\boldsymbol{\sigma}_m^{(t+1)}\}_{m=1}^{M}$. Then, with the updated IPC thresholds fixed, each BS $m$ solves its local rotation optimization problem to obtain the new rotation vector $\mathbf{z}_m^{(t+1)}$. The above two steps are repeated until the maximum AO iteration number $T_{\max}$ is reached or the algorithm converges.

 \begin{figure}[t]
	\centering
	\includegraphics[width=3in]{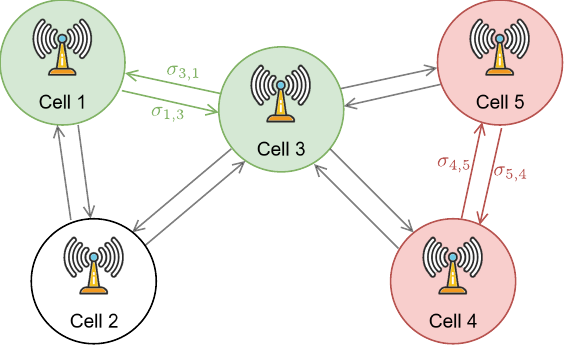} 
	\caption{Illustration of the  proposed IPC coordination iteration. A random maximal matching selects the two vertex-disjoint adjacent-cell pairs $(1,3)$ and $(4,5)$, highlighted in green and red, respectively. Since the selected pairs do not share any endpoint, the IPC thresholds $\{\sigma_{1,3},\sigma_{3,1}\}$ and $\{\sigma_{4,5},\sigma_{5,4}\}$ can be updated in parallel. All IPC thresholds associated with the remaining gray links are kept unchanged in the current IPC coordination iteration.}
	\label{IPC_update_map}
\end{figure}

\subsubsection{IPC Coordination}

At the $t$-th AO iteration, with the current 6DMA rotations $\mathbf{z}^{(t)}$ fixed, all $M$ cells coordinate their IPC thresholds in an edge-wise decentralized manner. 
Specifically, we adopt an iterative pairwise coordination procedure over the cell-adjacency graph. In each IPC coordination iteration, a set of vertex-disjoint adjacent-cell pairs is first selected, and then each selected pair updates its two associated IPC thresholds based on local rate evaluations. Let $\mathcal{G}=(\mathcal{V},\mathcal{E})$ denote the cell-adjacency graph, where $\mathcal{V}=\{1,\ldots,M\}$ and $(m,n)\in\mathcal{E}$ if cells $m$ and $n$ are adjacent. Let $T_{\mathrm{IPC}}$ denote the maximum number of IPC coordination iterations. At the $\tau$-th IPC coordination iteration, where $\tau=1,\ldots,T_{\mathrm{IPC}}$, instead of updating all adjacent-cell pairs simultaneously, we first construct a random maximal matching $\mathcal{L}^{(\tau)}\subseteq\mathcal{E}$. Since any two edges in $\mathcal{L}^{(\tau)}$ are vertex-disjoint, all selected pairs can be updated in parallel without conflicts, while the remaining edges stay unchanged in the current IPC coordination iteration. Fig.~\ref{IPC_update_map} illustrates one such IPC coordination iteration.

Let $\sigma_{m,n}^{(t,\tau)}$ denote the IPC threshold from cell $m$ to cell $n$ at the beginning of the $\tau$-th IPC coordination iteration within the $t$-th AO iteration. Accordingly, the IPC vector associated with BS $m$ is defined as
\begin{equation}
	\boldsymbol{\sigma}_m^{(t,\tau)}
	\triangleq
	\left[
	\{\sigma_{m,n}^{(t,\tau)}\}_{n\in\mathcal{A}_m},
	\{\sigma_{n,m}^{(t,\tau)}\}_{n\in\mathcal{A}_m}
	\right]^T
	\in \mathbb{R}_+^{2|\mathcal{A}_m|\times 1}.
	\label{eq:sigma_m_t_tau}
\end{equation}

The IPC coordination is initialized by setting $\boldsymbol{\sigma}_m^{(t,1)}=\boldsymbol{\sigma}_m^{(t)}$. For a selected pair $(m,n)\in\mathcal{L}^{(\tau)}$, only the two directed IPC thresholds $\sigma_{m,n}$ and $\sigma_{n,m}$ are updated, while all the other IPC thresholds are fixed at their current values. Let	$\boldsymbol{\sigma}_{m|n}^{(t,\tau)}(x,y)$
denote the vector obtained from the current IPC vector $\boldsymbol{\sigma}_m^{(t,\tau)}$ by replacing the entries $\sigma_{m,n}^{(t,\tau)}$ and $\sigma_{n,m}^{(t,\tau)}$ with $x$ and $y$, respectively, while keeping all other entries unchanged. Similarly, define $\boldsymbol{\sigma}_{n|m}^{(t,\tau)}(x,y)$ for BS $n$.
For a selected pair $(m,n)\in\mathcal{L}^{(\tau)}$, we consider the following pairwise coordination objective,
	\begin{align}
		\max_{x,y\in \mathbb{R}_{+}} \quad
		& \widehat{R}_m\!\left(
		\mathbf{z}_m^{(t)},
		\boldsymbol{\sigma}_{m|n}^{(t,\tau)}(x,y)
		\right)
		+
		\widehat{R}_n\!\left(
		\mathbf{z}_n^{(t)},
		\boldsymbol{\sigma}_{n|m}^{(t,\tau)}(x,y)
		\right).\label{IPC_coordination_ori_revise}
	\end{align}
 For notational convenience, define the corresponding pairwise SAA utility as
\begin{equation}
	\widehat{\Phi}_{m,n}^{(t,\tau)}(x,y)
	\triangleq
	\widehat{R}_m\!\left(
	\mathbf{z}_m^{(t)},
	\boldsymbol{\sigma}_{m|n}^{(t,\tau)}(x,y)
	\right)
	+
	\widehat{R}_n\!\left(
	\mathbf{z}_n^{(t)},
	\boldsymbol{\sigma}_{n|m}^{(t,\tau)}(x,y)
	\right).
	\label{eq:pairwise_saa_utility}
\end{equation}
Then, the pairwise coordination objective in \eqref{IPC_coordination_ori_revise} can be equivalently written as
\begin{equation}
	\max_{x,y\in\mathbb{R}_{+}}
	\widehat{\Phi}_{m,n}^{(t,\tau)}(x,y).
	\label{eq:pairwise_saa_utility_compact}
\end{equation}

Since directly solving \eqref{eq:pairwise_saa_utility_compact} may still incur a non-negligible search overhead, we adopt the following two-stage one-dimensional grid search for each selected pair $(m,n)\in\mathcal{L}^{(\tau)}$. Define the set of directed index pairs associated with the BS pair $(m,n)$ as
\begin{equation}
	\mathcal{P}_{m,n} \triangleq \{(m,n),(n,m)\}.
\end{equation}
For each $(i,j)\in\mathcal{P}_{m,n}$, let
\begin{align}
	\sigma_{i,j}^{\mathrm{lb},(t,\tau)}
	&\triangleq
	\max\!\left\{0.1\sigma_{i,j}^{(t,\tau)},\,\sigma_{\mathrm{fl}}\right\},\\
	\sigma_{i,j}^{\mathrm{ub},(t,\tau)}
	&\triangleq
	\max\!\left\{10\sigma_{i,j}^{(t,\tau)},\,\sigma_{i,j}^{\mathrm{lb},(t,\tau)}\right\},
\end{align}
where $\sigma_{\mathrm{fl}}>0$ is a small numerical floor. The logarithmic candidate set for $\sigma_{i,j}$ is given by
\begin{equation}
	\mathcal{C}_{i,j}^{(t,\tau)}
	\triangleq
	\left\{\sigma_{i,j}^{(t,\tau)}\right\}
	\cup
	\left\{
	\sigma_{i,j}^{\mathrm{lb},(t,\tau)}
	\left(
	\frac{\sigma_{i,j}^{\mathrm{ub},(t,\tau)}}
	{\sigma_{i,j}^{\mathrm{lb},(t,\tau)}}
	\right)^{\frac{r}{N_{\mathrm{g}}-1}}
	\right\}_{r=0}^{N_{\mathrm{g}}-1},
	\label{eq:ipc_grid_set}
\end{equation}
where $N_{\mathrm{g}}>1$ is a prescribed integer denoting the number of logarithmically spaced grid points generated between $\sigma_{i,j}^{\mathrm{lb},(t,\tau)}$ and $\sigma_{i,j}^{\mathrm{ub},(t,\tau)}$.

Starting from the current pair
$\big(\sigma_{m,n}^{(t,\tau)},\sigma_{n,m}^{(t,\tau)}\big)$, define
\begin{equation}
	\widehat{\Phi}_{m,n}^{\mathrm{cur},(t,\tau)}
	\triangleq
	\widehat{\Phi}_{m,n}^{(t,\tau)}\!\left(
	\sigma_{m,n}^{(t,\tau)},\sigma_{n,m}^{(t,\tau)}
	\right).
	\label{eq:ipc_pair_cur_value}
\end{equation}
The first one-dimensional search optimizes $\sigma_{m,n}$ while keeping $\sigma_{n,m}$ fixed, i.e.,
\begin{equation}
	\bar{\sigma}_{m,n}^{(t,\tau)}
	=
	\arg\max_{x\in\mathcal{C}_{m,n}^{(t,\tau)}}
	\widehat{\Phi}_{m,n}^{(t,\tau)}\!\left(
	x,\sigma_{n,m}^{(t,\tau)}
	\right),
	\label{eq:ipc_grid_search_mn}
\end{equation}
which can be efficiently implemented in parallel over all candidates in $\mathcal{C}_{m,n}^{(t,\tau)}$, since each candidate value of $x$ leads to an independent evaluation of $\widehat{\Phi}_{m,n}^{(t,\tau)}(x,\sigma_{n,m}^{(t,\tau)})$.  It is worth noting that this search only requires  limited pairwise information exchange between BSs $m$ and $n$, involving the candidate IPC threshold values and the corresponding scalar local utility values. 

Then, with $\bar{\sigma}_{m,n}^{(t,\tau)}$ fixed, the second one-dimensional search optimizes $\sigma_{n,m}$ similarly as
\begin{equation}
	\bar{\sigma}_{n,m}^{(t,\tau)}
	=
	\arg\max_{y\in\mathcal{C}_{n,m}^{(t,\tau)}}
	\widehat{\Phi}_{m,n}^{(t,\tau)}\!\left(
	\bar{\sigma}_{m,n}^{(t,\tau)},y
	\right).
	\label{eq:ipc_grid_search_nm}
\end{equation} 
After the two-stage search, the resulting pairwise utility is
\begin{equation}
	\widehat{\Phi}_{m,n}^{\mathrm{new},(t,\tau)}
	\triangleq
	\widehat{\Phi}_{m,n}^{(t,\tau)}\!\left(
	\bar{\sigma}_{m,n}^{(t,\tau)},
	\bar{\sigma}_{n,m}^{(t,\tau)}
	\right).
	\label{eq:ipc_pair_new_value}
\end{equation}

The candidate pair is accepted only if it provides a sufficiently positive pairwise gain, namely,
\begin{equation}
	\widehat{\Phi}_{m,n}^{\mathrm{new},(t,\tau)}
	>
	\widehat{\Phi}_{m,n}^{\mathrm{cur},(t,\tau)}
	+
	\epsilon_{\mathrm{tol}},
	\label{eq:IPC_grid_accept}
\end{equation}
where $\epsilon_{\mathrm{tol}}>0$ is a small numerical tolerance. If \eqref{eq:IPC_grid_accept} holds, the accepted IPC update for pair $(m,n)$ is
\begin{equation}
	\sigma_{m,n}^{(t,\tau+1)}
	=
	\bar{\sigma}_{m,n}^{(t,\tau)},
	\qquad
	\sigma_{n,m}^{(t,\tau+1)}
	=
	\bar{\sigma}_{n,m}^{(t,\tau)}.
	\label{eq:IPC_pair_output_grid}
\end{equation}

After all pairs in $\mathcal{L}^{(\tau)}$ complete the above procedure, the IPC thresholds on the selected edges are updated simultaneously according to the accepted grid-search results.  The above random-matching-based IPC coordination is repeated for $T_{\mathrm{IPC}}$ IPC coordination iterations. After these iterations, the resulting IPC vector of each BS $m$ is taken as the IPC output of the $t$-th AO iteration, i.e.,
\begin{equation}
	\boldsymbol{\sigma}_m^{(t+1)}
	=
	\boldsymbol{\sigma}_m^{(t,T_{\mathrm{IPC}}+1)},\quad \forall m\in\mathcal{M}.
\end{equation}

\subsubsection{Rotation Optimization}

At the same AO iteration, given the updated IPC thresholds $\boldsymbol{\sigma}_m^{(t+1)}$, the long-term rotation variables of BS $m$ are updated by solving
\begin{equation}
	\max_{\mathbf{z}_m \in \mc{Z}_m}
	\ \widehat{R}_m\!\left(\mathbf{z}_m,\boldsymbol{\sigma}_m^{(t+1)}\right).
	\label{eq:rot_opt_prob}
\end{equation}

Since problem \eqref{eq:rot_opt_prob} is generally nonconvex with respect to the angular variables, we employ PSO as a practical local solver \cite{KennedyEberhart1995PSO}.
For subsequent complexity analysis, let $N_{\mathrm{sw}}$ and $T_{\mathrm{PSO}}$ denote the swarm size and the maximum number of PSO iterations, respectively.
 Specifically, we equivalently minimize the negative average cell rate
\begin{equation}
f_m(\mathbf{z}_m)
=
-\widehat{R}_m\!\left(\mathbf{z}_m,\boldsymbol{\sigma}_m^{(t+1)}\right).
	\label{eq:pso_obj_rot}
\end{equation}
To incorporate the constraint $\mathbf{z}_m \in \mc{Z}_m$ into the PSO search, we adopt a penalty-based formulation and consider
\begin{equation}
\tilde{f}_m(\mathbf{z}_m)
=
-\widehat{R}_m\!\left(\mathbf{z}_m,\boldsymbol{\sigma}_m^{(t+1)}\right)
+
\kappa \Psi(\mathbf{z}_m),
	\label{eq:pso_penalty}
\end{equation}
where $\Psi(\mathbf{z}_m)$ is a nonnegative penalty function characterizing the violation of the feasible set $\mc{Z}_m$, and $\kappa>0$ is the corresponding penalty coefficient. By construction, $\Psi(\mathbf{z}_m)=0$ holds for any feasible $\mathbf{z}_m \in \mc{Z}_m$, while $\Psi(\mathbf{z}_m)>0$ otherwise.
Starting from the current feasible rotation vector, the swarm iteratively updates a set of candidate solutions, and the best feasible particle returned by the solver is taken as the updated rotation solution $\mathbf{z}_m^{(t+1)}$. 
 
 The overall decentralized long-term AO procedure is summarized in Algorithm~\ref{alg:long_term_ao}.

\begin{algorithm}[t!]
	\caption{Decentralized Long-Term AO Optimization}
	\label{alg:long_term_ao}
	\begin{algorithmic}[1]
\STATE \textbf{Input} $\{\mathbf{z}_m^{(0)}\}_{m\in\mathcal{M}}$, $\{\bm \sigma_{m}^{(0)}\}_{m\in\mathcal{M}}$, $\mathcal{G}=(\mathcal{V},\mathcal{E})$,  $T_{\max}$, $T_{\mathrm{IPC}}$, $N_{\mathrm{g}}$, $\sigma_{\mathrm{fl}}$, $\epsilon_{\mathrm{tol}}$, $N_{\mathrm{sw}}$, $T_{\mathrm{PSO}}$, and $\kappa$.
		
		\STATE Set $t \gets 0$.
		
		\REPEAT
		\STATE With $\{\mathbf{z}_m^{(t)}\}_{m=1}^{M}$ fixed, initialize the IPC coordination by setting $\boldsymbol{\sigma}_m^{(t,1)}=\boldsymbol{\sigma}_m^{(t)}$, $\forall m\in\mathcal{M}$.
		
		\FOR{$\tau=1,\ldots,T_{\mathrm{IPC}}$}
		\STATE Randomly construct a maximal matching $\mathcal{L}^{(\tau)}\subseteq\mathcal{E}$.
		
		\FOR{each selected pair $(m,n)\in\mathcal{L}^{(\tau)}$ \textbf{in parallel}}
		\STATE Update the corresponding IPC thresholds by following the grid-search procedure in \eqref{eq:ipc_grid_set}--\eqref{eq:IPC_pair_output_grid}.
		\ENDFOR
		
		\ENDFOR
		
		\STATE Set $\boldsymbol{\sigma}_m^{(t+1)}=\boldsymbol{\sigma}_m^{(t,T_{\mathrm{IPC}}+1)}$, $\forall m\in\mathcal{M}$.
		
		\FOR{each BS $m=1,\ldots,M$ \textbf{in parallel}}
		\STATE Starting from $\mathbf{z}_m^{(t)}$, solve \eqref{eq:rot_opt_prob} with $\boldsymbol{\sigma}_m^{(t+1)}$ fixed by using the PSO-based update in \eqref{eq:pso_obj_rot} and \eqref{eq:pso_penalty}, and obtain $\mathbf{z}_m^{(t+1)}$.
		\ENDFOR
		
		\STATE Apply  $\{\boldsymbol{\sigma}_m^{(t+1)}\}_{m=1}^{M}$ and $\{\mathbf{z}_m^{(t+1)}\}_{m=1}^{M}$ at all BSs for the subsequent transmission blocks.
		
		\STATE $t \gets t+1$.
		
		\UNTIL{$t = T_{\max}$}
		
		\STATE \textbf{Output} Updated IPC thresholds $\{\boldsymbol{\sigma}_m^{(t)}\}_{m=1}^{M}$ and rotation vectors $\{\mathbf{z}_m^{(t)}\}_{m=1}^{M}$.
	\end{algorithmic}
\end{algorithm}

\subsection{Computational Complexity}
\label{subsec:complexity_dist}

We evaluate the computational complexity of the proposed decentralized two-timescale algorithm at BS $m$. Let $T_{\mathrm{dual}}$ denote the number of inner dual updates for the IPC multipliers, and  $T_{\mu}$ denote the number of bisection steps for satisfying the power constraint. Define the effective short-term iteration count as
$
	T_{\mathrm{WMMSE}}'
	\triangleq
	T_{\mathrm{WMMSE}}\,T_{\mathrm{dual}}\,T_{\mu}.
	\label{eq:Twmmse_eff}
$

\textit{Short-term precoder update:}
For a fixed rotation vector $\mbf z_m$ and one channel realization, the short-term solver is dominated by the inversion of an $NB\times NB$ matrix in the closed-form precoder update. Therefore, the per-sample short-term optimization complexity at BS $m$ scales as
$
	\mc O\!\left(
	T_{\mathrm{WMMSE}}'(NB)^3
	\right)
	\label{eq:dist_shortterm_complexity}
$.

\textit{IPC coordination:}
At each IPC coordination iteration, due to the random maximal matching, BS $m$ participates in at most one selected adjacent-cell pair. For each selected pair, the two-stage one-dimensional grid search involves $\mathcal{O}(N_{\mathrm{g}})$ candidate utility evaluations, which can be parallelized over the grid points. Thus, under parallel candidate evaluation, its effective evaluation complexity is $\mathcal{O}(1)$ with respect to $N_{\mathrm{g}}$. 
Each IPC utility evaluation is obtained from $S$ independent short-term problem realizations. Hence, the IPC-coordination complexity at BS $m$ in one AO iteration scales as
\begin{equation}
	\mathcal{O}\!\left(
	T_{\mathrm{IPC}}\,
	S\,
	T_{\mathrm{WMMSE}}'(NB)^3
	\right).
	\label{eq:dist_ipc_complexity}
\end{equation}

\textit{Rotation optimization:}
At each PSO iteration, each particle requires one long-term objective evaluation. Since the penalty term associated with the feasible set $\mc Z_m$ is of lower complexity than the average-rate evaluation, the per-particle cost is dominated by computing the average cell rate. Each such evaluation is again obtained by averaging over $S$ independent short-term problem realizations. Therefore, the rotation-update complexity at BS $m$ in one AO iteration scales as
\begin{equation}
	\mc O\!\left(
	N_{\mathrm{sw}}\,
	T_{\mathrm{PSO}}\,
	S\,
	T_{\mathrm{WMMSE}}'(NB)^3
	\right).
	\label{eq:dist_rot_complexity}
\end{equation}

\textit{Overall long-term optimization complexity:}
Since the IPC coordination and the PSO-based rotation optimization are both executed once in each AO iteration, we define the effective long-term evaluation count over all AO iterations as
$
	T_{\mathrm{PSO}}'
	\triangleq
	T_{\max}\!\left(
	T_{\mathrm{IPC}}
	+
	N_{\mathrm{sw}}T_{\mathrm{PSO}}
	\right).
	\label{eq:Tlong_eff}
$
Then, by combining \eqref{eq:dist_ipc_complexity} and \eqref{eq:dist_rot_complexity}, the overall long-term optimization complexity at BS $m$ scales as
\begin{equation}
	\mc O\!\left(
	T_{\mathrm{PSO}}'\,
	S\,
	T_{\mathrm{WMMSE}}'(NB)^3
	\right).
	\label{eq:dist_total_complexity_perBS}
\end{equation}

An important observation from \eqref{eq:dist_total_complexity_perBS} is that the dominant per-BS computational cost of the proposed decentralized method does not explicitly depend on the number of cells $M$. This is because the short-term precoder update at each BS only requires matrix inversions of size $NB$, while the local PSO-based rotation update is carried out over the $2B$-dimensional local rotation vector $\mathbf{z}_m$. In contrast, the centralized offline benchmark in Section~III optimizes the global rotation vector of dimension $2BM$, and its long-term search generally requires a larger search budget as $M$ increases. Therefore, under parallel implementation across BSs, the proposed decentralized method admits substantially better scalability than the centralized benchmark as the network size increases.

\section{Simulation Results}\label{sec:sim}

This section provides simulation results to validate the proposed decentralized long-term optimization framework for IPC coordination and 6DMA rotation optimization. In the simulation, we consider a three-cell downlink network with $M=3$ BSs. Each BS is equipped with $B=3$ 6DMA surfaces and each surface contains $N=4$ antennas, so the total number of transmit antennas per BS is $NB = 12$. The carrier wavelength is $\lambda=0.125$  meter (m), and the four antennas on each square 6DMA surface are located at $[0,\pm \frac{\lambda}{4},\pm \frac{\lambda}{4}]$ in its local coordinate system. The channel coefficient $\xi_{m,k}$ is modeled as a circularly symmetric complex Gaussian random variable with zero mean, i.e.,
\begin{equation}
	\xi_{m,k} \sim \mathcal{CN}\!\left(0,\beta(d_{m,k})\right), \quad m\in\mc M,~k\in\mc K,
	\label{eq:xi_k_distribution}
\end{equation}
where $\beta(d_{m,k})$ denotes the large-scale fading coefficient associated with the propagation distance $d_{m,k}$ from the $m$-th 6DMA-BS to the $k$-th user, given by
\begin{equation}
	\beta(d_{m,k})=\left(\frac{\lambda}{4\pi}\right)^2 d_{m,k}^{-\eta},
	\label{eq:large_scale_fading_k}
\end{equation}
with path loss exponent $\eta=3.5$. The antenna gain pattern in (\ref{antenna_gain}) follows the 3GPP standard \cite{3gpp2017}.

\begin{figure}[!t]
	\centering
	\includegraphics[width=3in]{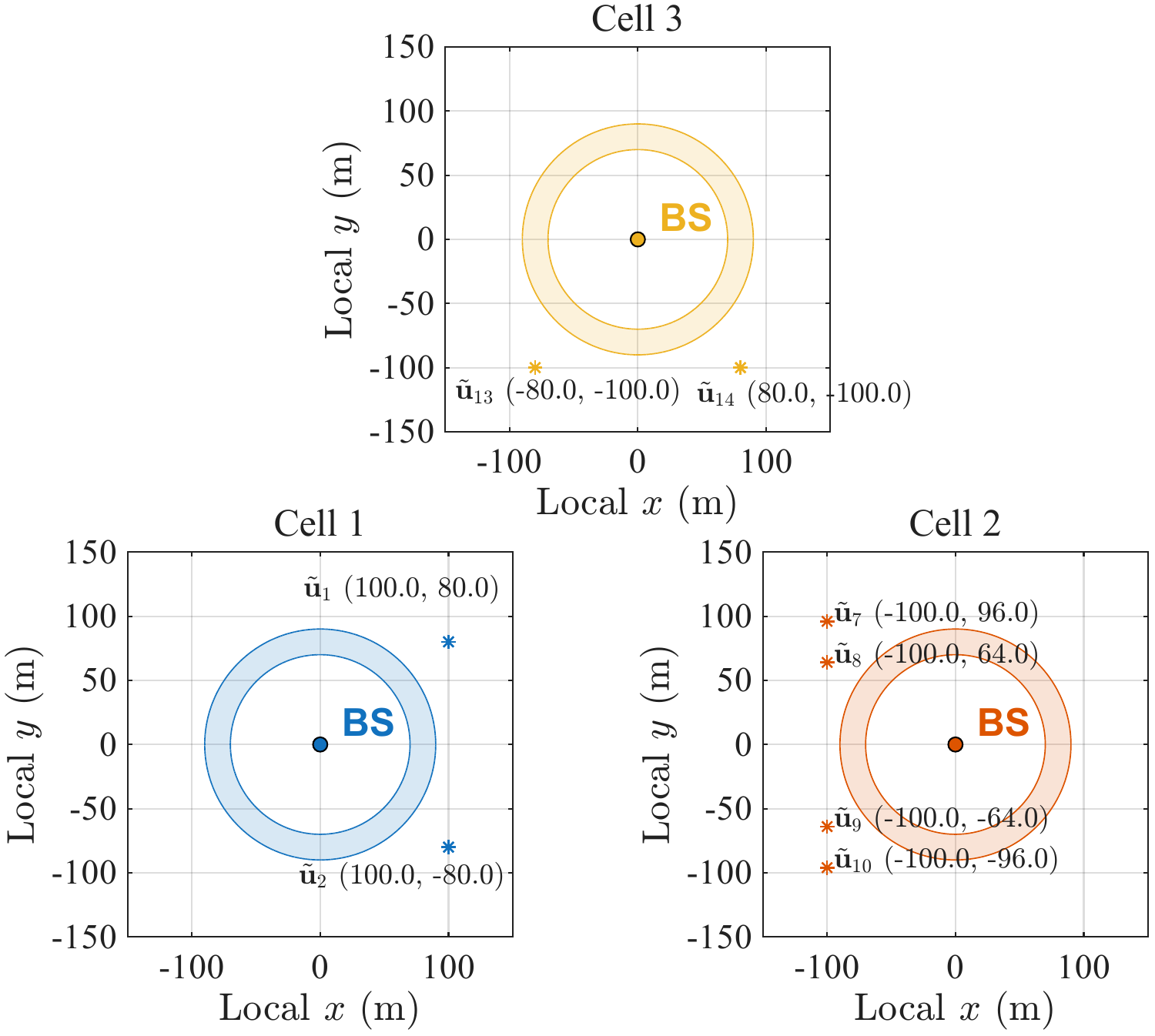}
	\caption{Local user distributions in the three cells. In each subfigure, the BS is located at the origin of the local coordinate system. The shaded annulus represents the region of regular users, while the marked users denote the fixed edge users  to create asymmetric spatial distributions across the cells.}
	\label{fig:user_loc}
\end{figure}

To flexibly generate different levels of interference  while keeping the in-cell user geometry unchanged, we adopt a fixed local user geometry with a movable cell layout. Specifically, for each cell $m$, the user locations are specified in the BS-centered local coordinate system as $\{\tilde{\mathbf{u}}_{k}\in\mathbb{R}^{3\times 1}\}_{k\in\mathcal{K}_m}$, which remain fixed throughout the simulation, as illustrated in Fig.~\ref{fig:user_loc}. The corresponding global user positions are given by
\begin{equation}
	\mathbf{x}_k=\mathbf{o}_m+\tilde{\mathbf{u}}_{k},\quad \forall m\in\mathcal M,\ \forall k\in\mathcal K_m, \label{user_loc_map}
\end{equation}
where $\mathbf{o}_m$ denotes the global location of BS $m$. 
For each cell $m$, the user set is partitioned as $\mathcal{K}_m=\mathcal{K}_m^{\mathrm{reg}}\cup\mathcal{K}_m^{\mathrm{edge}}$, where $\mathcal{K}_m^{\mathrm{reg}}$ and $\mathcal{K}_m^{\mathrm{edge}}$ denote the regular users and the edge users, respectively. We assign different service priorities to these two user types by setting the user-rate weight to $1$ for edge users and $0.5$ for regular users. In each cell, the regular users are placed in the local horizontal plane around the serving BS within an annular region with inner radius $70$~m and outer radius $90$~m. Specifically, we set $\mathcal{K}_1^{\mathrm{reg}}=\{3,4,5,6\}$, $\mathcal{K}_2^{\mathrm{reg}}=\{11,12\}$, and $\mathcal{K}_3^{\mathrm{reg}}=\{15,16,17,18\}$. The edge-user locations are deterministic and fixed in the local coordinate system throughout the simulation. Specifically, we set $\mathcal{K}_1^{\mathrm{edge}}=\{1,2\}$, $\mathcal{K}_2^{\mathrm{edge}}=\{7,8,9,10\}$, and $\mathcal{K}_3^{\mathrm{edge}}=\{13,14\}$, whose local coordinates are illustrated in Fig.~\ref{fig:user_loc}. These local user geometries are then mapped to the global coordinates via \eqref{user_loc_map}, which enables flexible control of the ICI level by adjusting $\{\mathbf{o}_m\}$ while keeping the in-cell user geometry unchanged. 

The maximum transmit power is $P_{\max}=50~\mathrm{dBm}$, i.e., $100~\mathrm{W}$, and the additive Gaussian noise power is $\sigma^2=-80~\mathrm{dBm}$, i.e., $10^{-8}~\mathrm{mW}$.
For the IPC coordination step, we define the adjacent-cell edge set as $\{(1,2),(1,3),(2,3)\}$, which means that the three cells form a fully connected adjacency graph, i.e., every pair of cells is adjacent.  
Equivalently, the neighbor sets are explicitly given by
\begin{equation}
	\mathcal{A}_1=\{2,3\},\quad \mathcal{A}_2=\{1,3\},\quad \mathcal{A}_3=\{1,2\}.
\end{equation}
In each AO iteration, the IPC thresholds are first updated by the proposed random-matching-based two-stage one-dimensional grid-search rule. The numerical floor, the grid size, the acceptance tolerance, and the maximum number of IPC coordination iterations are set to $\sigma_{\mathrm{fl}}=10^{-9}$, $N_{\mathrm{g}}=7$, $\epsilon_{\mathrm{tol}}=10^{-4}$, and $T_{\mathrm{IPC}}=5$, respectively. For the PSO-based rotation update, the swarm size and the maximum number of PSO iterations are set to $N_{\mathrm{sw}}=12$ and $T_{\mathrm{PSO}}=60$, respectively. The penalty coefficient in \eqref{eq:pso_penalty} is set to $\kappa=10^{3}$.

For the rotation constraints, the azimuth angles are wrapped to $(0,2\pi]$, and the elevation angle of each 6DMA surface is restricted to the downward tilt range $[0^\circ,60^\circ]$. The minimum angular separation between the surface normals is set to $\Delta\phi=5^\circ$. The SAA approximation reuses the same set of $S=100$ channel samples for all long-term objective evaluations. Finally, the maximum number of AO iterations in Algorithm~\ref{alg:long_term_ao} is set to $T_{\max}=5$.

To demonstrate the performance of our proposed protocol under different ICI conditions, we consider two setups with different cell layouts, while keeping the local distribution of users in each cell unchanged. Setup~1 corresponds to a more interference-limited geometry with stronger ICI, whereas Setup~2 corresponds to a medium-ICI condition.
Unless otherwise specified, all coordinates in the following setups are in meters.
\begin{enumerate}
	\item {High-ICI setup:} The BS locations are set as 
	$\mathbf{o}_1=[0,0,10]^T$, $\mathbf{o}_2=[300,0,10]^T$, and 
	$\mathbf{o}_3=[150,280,10]^T$. In this setup, the BSs are placed relatively close to each other, and thus the ICI coupling is strong.
	
	\item {Medium-ICI setup:} The BS locations are set as 
	$\mathbf{o}_1=[0,0,10]^T$, $\mathbf{o}_2=[400,0,10]^T$, and 
	$\mathbf{o}_3=[200,380,10]^T$. In this setup, the BSs are more widely separated, and thus the ICI coupling is weaker than that in the high-ICI setup.
\end{enumerate}
For performance comparison, the following benchmark schemes are considered. Unless otherwise specified, the reported rates of all schemes are evaluated by using the actual SINR in \eqref{eq:SINR_inst}. For the schemes with fixed 6DMA rotations, the rotations are fixed at the uniformly distributed initial configuration.
\begin{enumerate}
	\item Rotation-only benchmark: The short-term precoders are obtained by solving the local WMMSE problem without IPC constraints. Accordingly, the 6DMA rotations are  optimized without accounting for ICI.
	
	\item IPC-coordination-only benchmark: The short-term precoders are solved based on our proposed coordination scheme, while the 6DMA rotations are fixed.

	\item Centralized upper bound: Both the short-term precoders and the 6DMA rotations are  optimized in a centralized manner as discussed in Section~III.
	
	\item Centralized fixed-position array (FPA) benchmark: The short-term precoding is obtained by the centralized method, while the 6DMA rotations are fixed.
	
	\item FPA benchmark: The short-term precoders are obtained by solving the local WMMSE problem without IPC constraints, and the 6DMA rotations are also fixed.
\end{enumerate}

\begin{figure}[!t]
	\centering
	\includegraphics[width=3in]{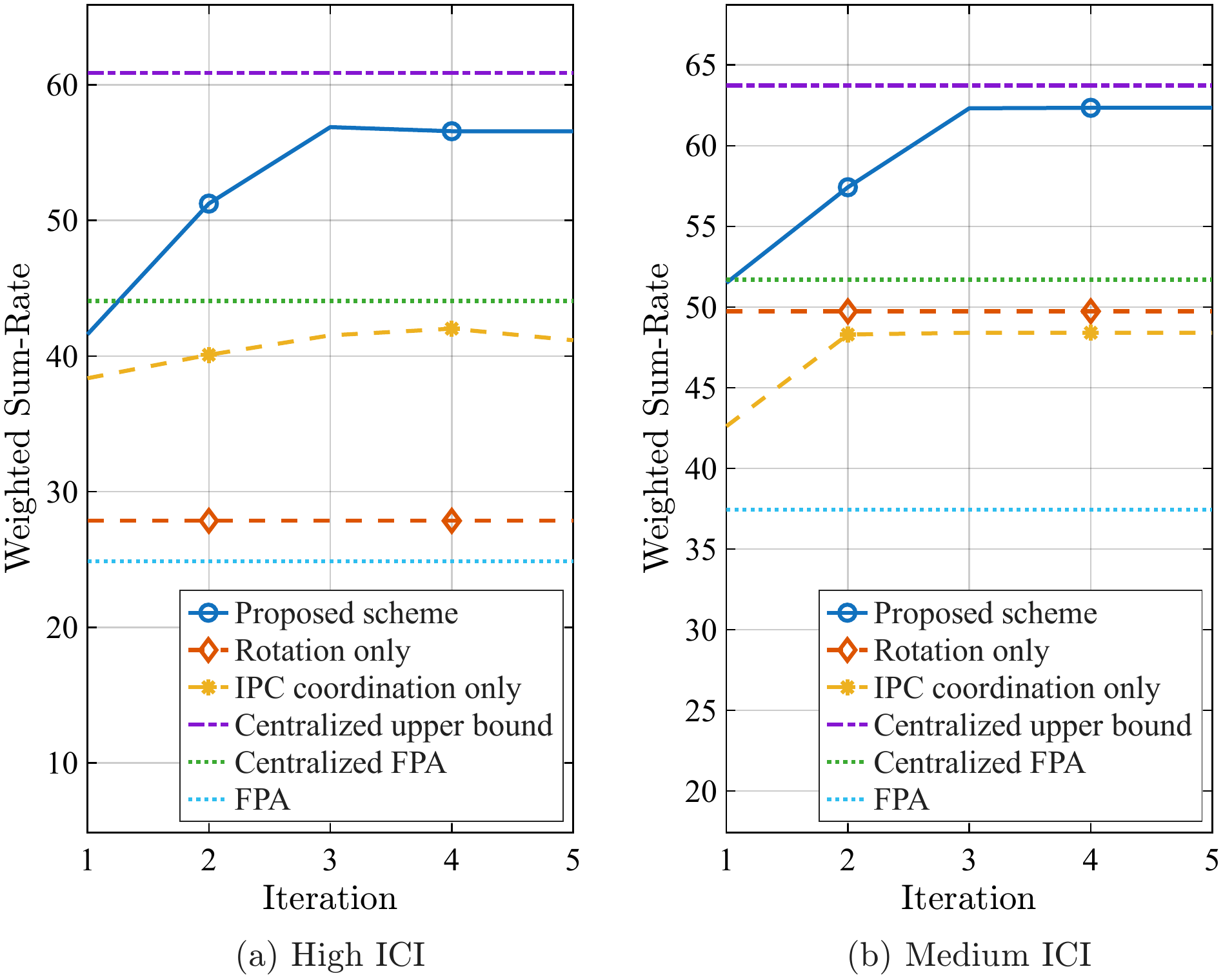}
\caption{Weighted sum-rate versus the long-term iteration index under different ICI levels. Fig.~\ref{fig:conv_logsumrate}(a) and Fig.~\ref{fig:conv_logsumrate}(b) correspond to the high-ICI and medium-ICI setups, respectively.}
	\label{fig:conv_logsumrate}
\end{figure}

First, Fig.~\ref{fig:conv_logsumrate} shows the weighted sum-rate versus the long-term iteration index under the high-ICI and medium-ICI setups. In both setups, the proposed scheme improves the rate performance  rapidly in the first few iterations and then gradually converges,  consistently outperforms all practical benchmark schemes, and remains below but close to the centralized upper bound in both setups, thereby corroborating the effectiveness of the proposed decentralized two-timescale design. Moreover, the results reveal a clear performance dependence of the benchmark schemes on the interference regime. In the high-ICI setup, the rotation-only benchmark suffers from a severe performance loss, which shows that simply enhancing the desired-signal power through array rotation is far from sufficient when ICI is strong. In contrast, in the medium-ICI setup, the rotation-only benchmark becomes much more effective, since the weaker ICI coupling allows desired-signal enhancement to play a more significant role.  
Another important observation is that the gap between the centralized upper bound and the IPC-coordination-only benchmark is smaller in the high-ICI setup than in the medium-ICI setup. Combined with the poor performance of the rotation-only benchmark under high ICI, this observation indicates that interference suppression is more critical than desired-signal enhancement in strongly interference-limited networks. By comparison, when the ICI coupling is moderate, further improving the user-link gain through rotation becomes increasingly important, which enlarges the performance gap between the IPC-coordination-only benchmark and the centralized upper bound.

\begin{figure}[!t]
	\centering
	\includegraphics[width=3in]{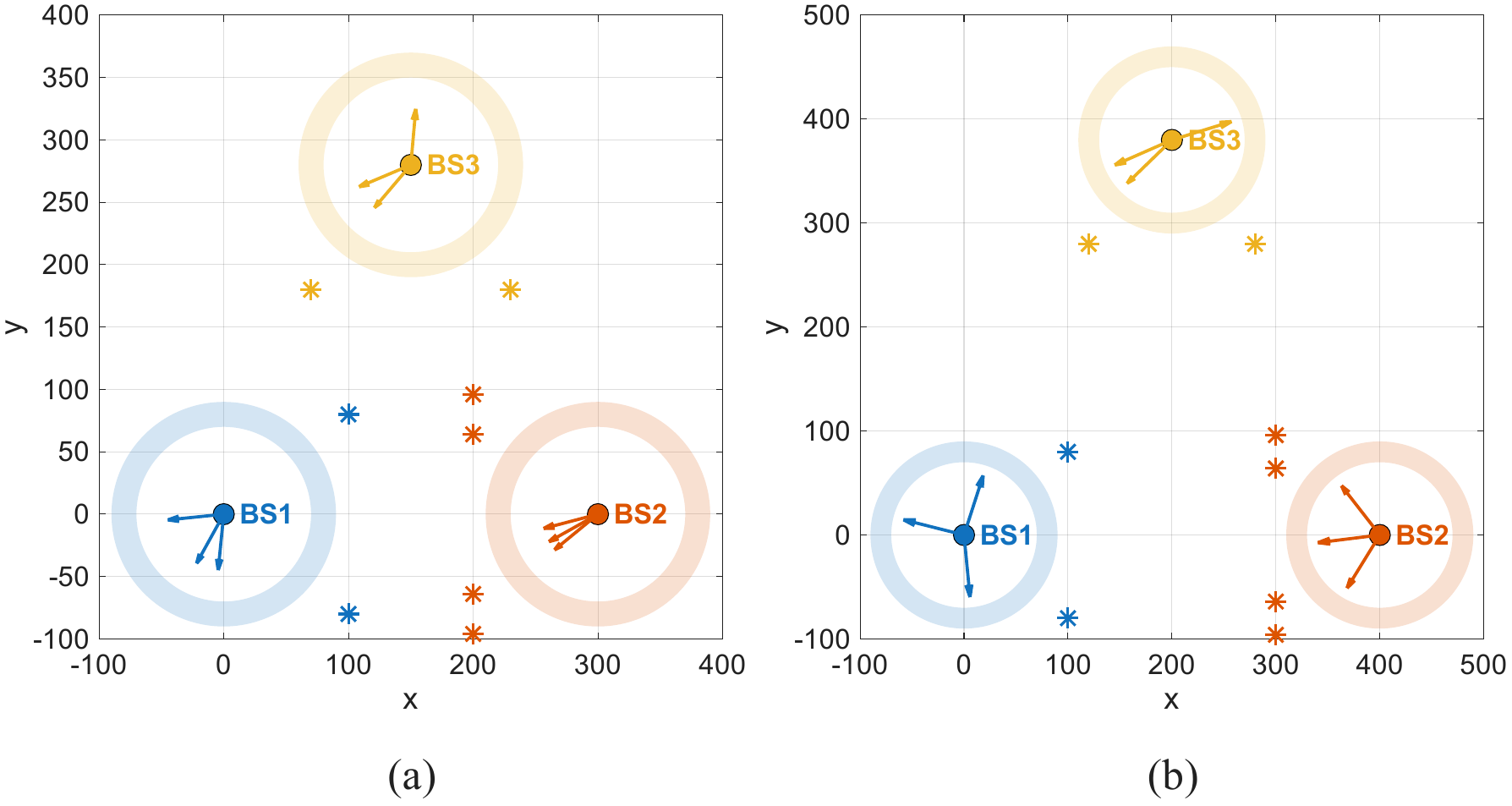}
	\caption{Illustration of two network geometries with different ICI levels. Fig.~\ref{fig:ICI_levels}(a) corresponds to a high-ICI setting with smaller BS separation, while Fig.~\ref{fig:ICI_levels}(b) corresponds to a medium-ICI setting with larger BS separation.}
	\label{fig:ICI_levels}
\end{figure}

Next, we compare two representative network geometries with different ICI levels, as illustrated in Fig.~\ref{fig:ICI_levels}. In the high-ICI setup shown in Fig.~\ref{fig:ICI_levels}(a), the BSs are placed closer to each other, and the resulting strong ICI coupling makes interference management a major concern in the long-term rotation design. Accordingly, the optimized 6DMA surface directions reflect a clear compromise between serving the local users and limiting leakage toward neighboring cells. In the medium-ICI setup shown in Fig.~\ref{fig:ICI_levels}(b), the BS separation becomes larger and the ICI coupling is weaker. The optimized rotations therefore place more emphasis on matching the local in-cell user distribution and enhancing the desired array gain, since the need for  interference suppression is less pronounced. 

\begin{figure}[!t]
	\centering
	\includegraphics[width=3in]{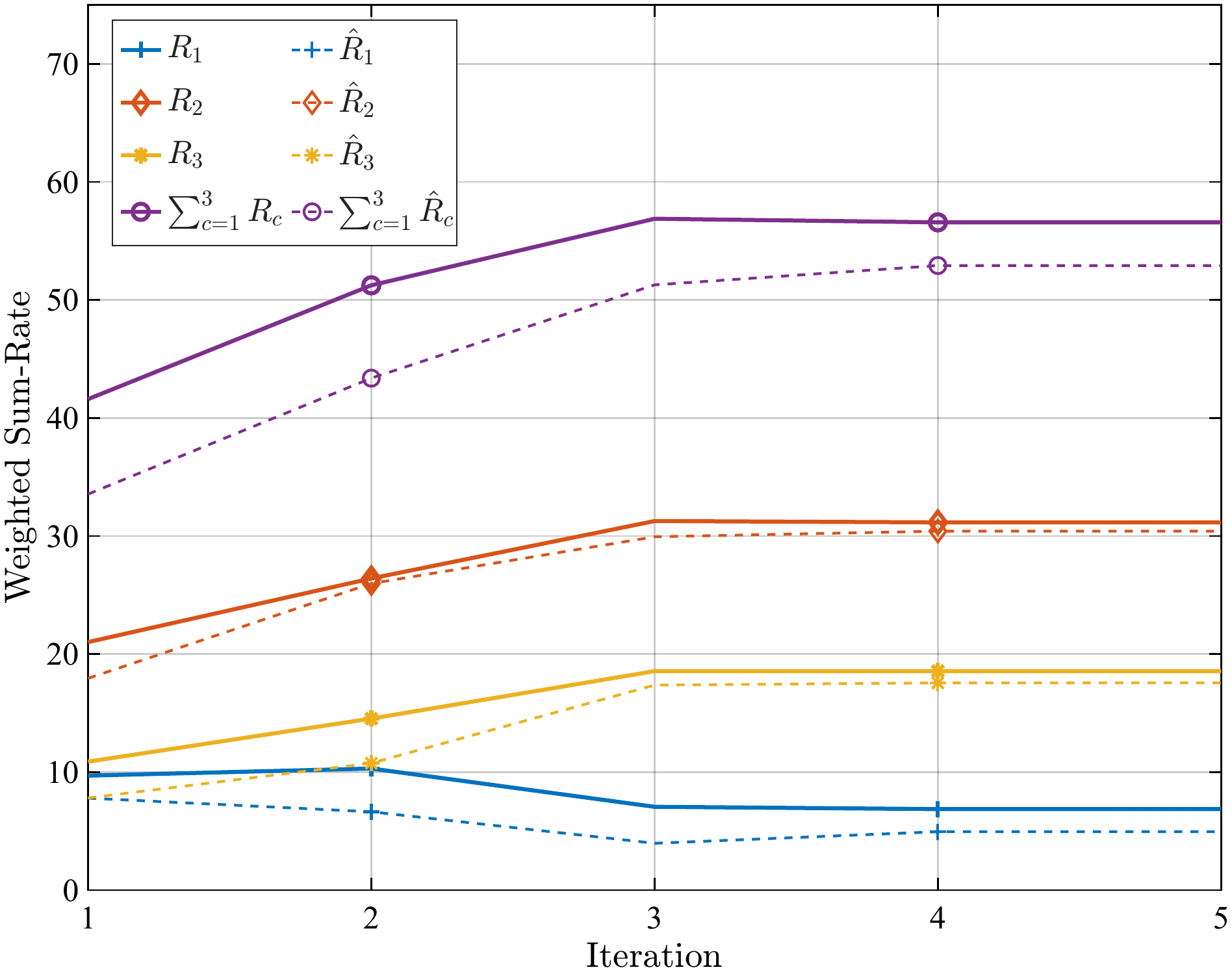}
\caption{Per-cell weighted rates and network weighted sum-rate versus optimization iteration for the proposed scheme. }
	\label{fig:cellrate_sumrate}
\end{figure}

Next, Fig.~\ref{fig:cellrate_sumrate} depicts the per-cell weighted rates and the network weighted sum-rate achieved by the proposed scheme over the iterations in the high-ICI case. In this figure, the solid curves represent the actual weighted rates obtained by substituting the updated rotation vectors into the original objective, while the dashed curves correspond to the surrogate rate using the lower-bounded SINR in \eqref{SINR_bound}  for the long-term optimization. It is observed that both the actual network weighted sum-rate and its surrogate counterpart increase rapidly in the first few iterations and then enter a stable regime with only mild fluctuations. Moreover, the surrogate sum-rate curve consistently remains below the actual one and follows a similar overall trend, which confirms that the adopted SINR lower bound provides a meaningful surrogate for guiding the 6DMA surface rotation update. The per-cell curves further reveal a heterogeneous rate evolution across the cells. In particular, the weighted rate of Cell~1 decreases gradually over the iterations, whereas those  of Cell~2 and Cell~3 increase significantly and become the main contributors to  the overall network gain. These observations indicate that the proposed long-term rotation design and short-term precoding adaptation jointly reshape the ICI distribution and reallocate antenna resources across cells so as to improve the overall weighted sum-rate. 

\begin{figure}[!t]
	\centering
	\includegraphics[width=3in]{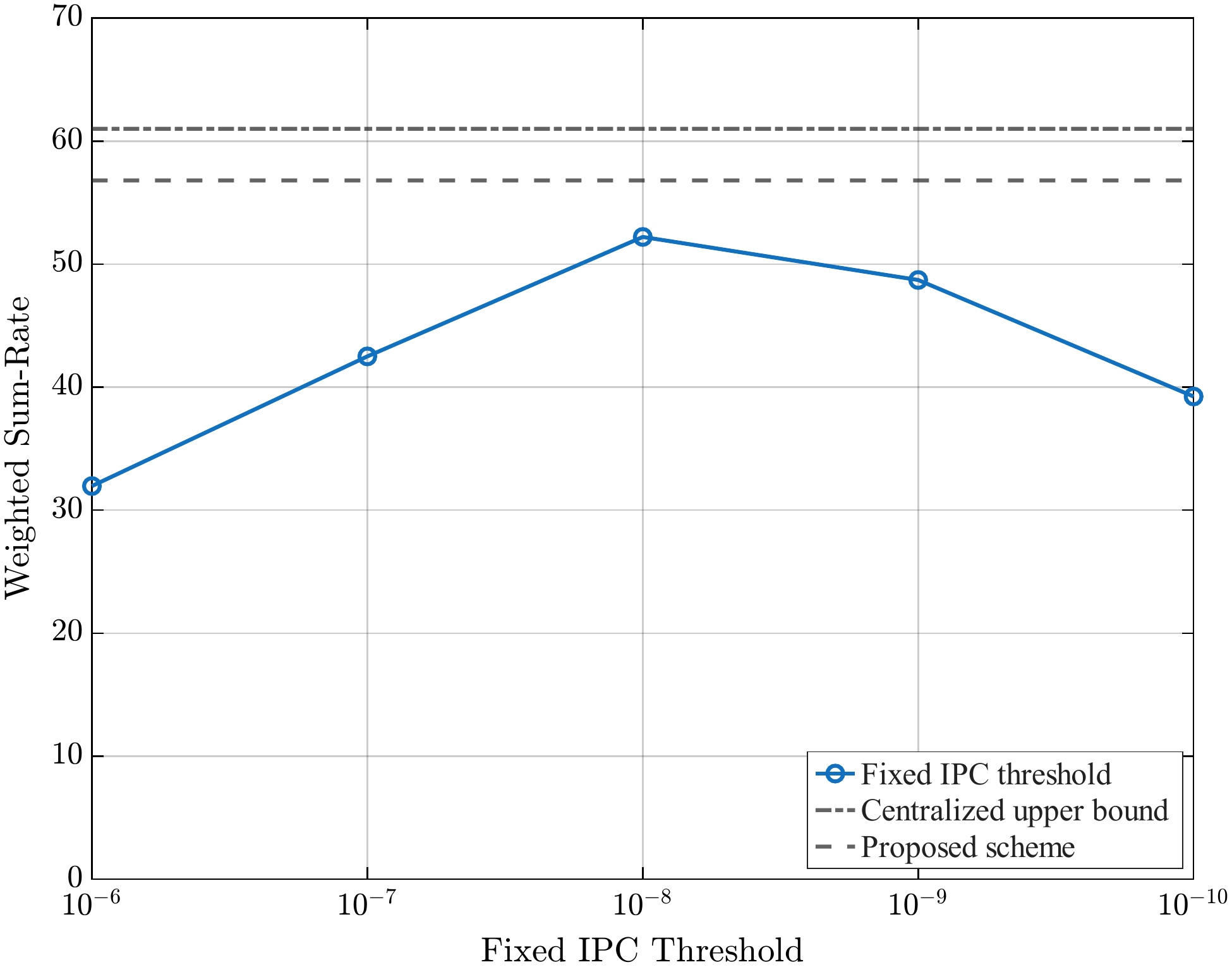}
	\caption{Weighted sum-rate versus a fixed IPC threshold. }
	\label{fig:fixed_IPC}
\end{figure}

Furthermore, Fig.~\ref{fig:fixed_IPC} evaluates the impact of using a fixed IPC threshold in the high-ICI setup. In this simulation, the rotation vectors are still optimized, whereas the proposed IPC coordination is disabled and all inter-cell IPC thresholds are set to the same constant value for every cell pair. As shown in Fig.~\ref{fig:fixed_IPC}, the achieved weighted sum-rate is highly sensitive to the selected IPC threshold and exhibits a clear non-monotonic behavior. When the threshold is too loose, such as $10^{-6}$, the interference constraint becomes insufficiently restrictive, allowing excessive inter-cell leakage and resulting in  a low weighted sum-rate. While as the threshold is reduced, the performance improves substantially and reaches its maximum value around $10^{-8}$. However, if the threshold is made overly stringent, for instance $10^{-10}$, the feasible precoding and rotation design becomes too conservative, which again causes a significant rate loss. Even at its best operating point, the fixed-threshold scheme remains below both the proposed scheme and the centralized upper bound, indicating that a single IPC threshold cannot flexibly match heterogeneous interference conditions across different cells and links. Nevertheless, if a well-predefined fixed IPC threshold is available, the fixed-threshold scheme can still serve as a low-complexity alternative in resource-limited scenarios, since it avoids iterative IPC coordination and reduces signaling overhead.

\begin{figure}[!t]
	\centering
	\includegraphics[width=3in]{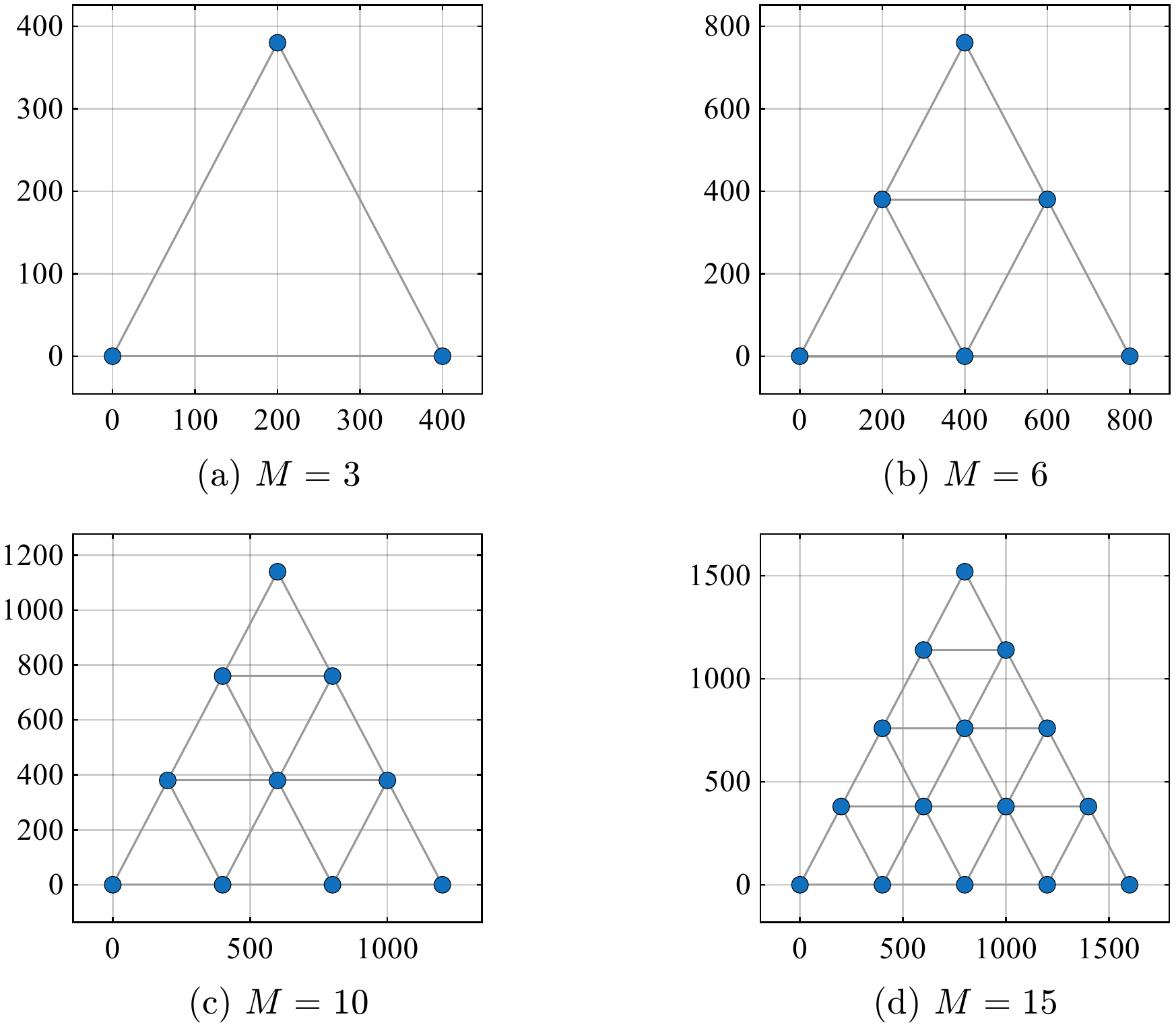}
\caption{Simulation setup for the scalability study under different network sizes. The BS topology is generated on a triangular lattice with $M=3$, $6$, $10$, and $15$ cells, where each node denotes one BS and each edge indicates an adjacency relation between two neighboring cells.}
	\label{fig:scalability_setup}
\end{figure}

To further examine the scalability of the proposed decentralized long-term optimization framework, we finally consider larger BS topologies with different network sizes. The purpose of this simulation is to investigate whether the proposed IPC coordination and local 6DMA rotation updates can still maintain good performance when the number of cells increases and the inter-cell adjacency graph becomes larger. In particular, we would like to assess how closely the proposed decentralized design can approach the corresponding centralized upper bound as the network expands. For this scalability study, the BS topology is generated on a triangular lattice with $M=3$, $6$, $10$, and $15$, as illustrated in Fig.~\ref{fig:scalability_setup}. Starting from the original three-cell geometry, larger networks are constructed by adding new rows of BSs on the same lattice, where the edge length of the smallest equilateral triangle is fixed to $400 \,\mathrm{m}$. To isolate the effect of network size from that of user-layout variations, we assume that all cells share the same local user distribution. Specifically, each cell contains six regular users and no edge user, where the regular users are generated in the same annular region as described in Fig.~\ref{fig:user_loc}. Unless otherwise specified, the remaining simulation parameters are the same as those used in the previous simulations.

\begin{figure}[!t]
	\centering
	\includegraphics[width=3in]{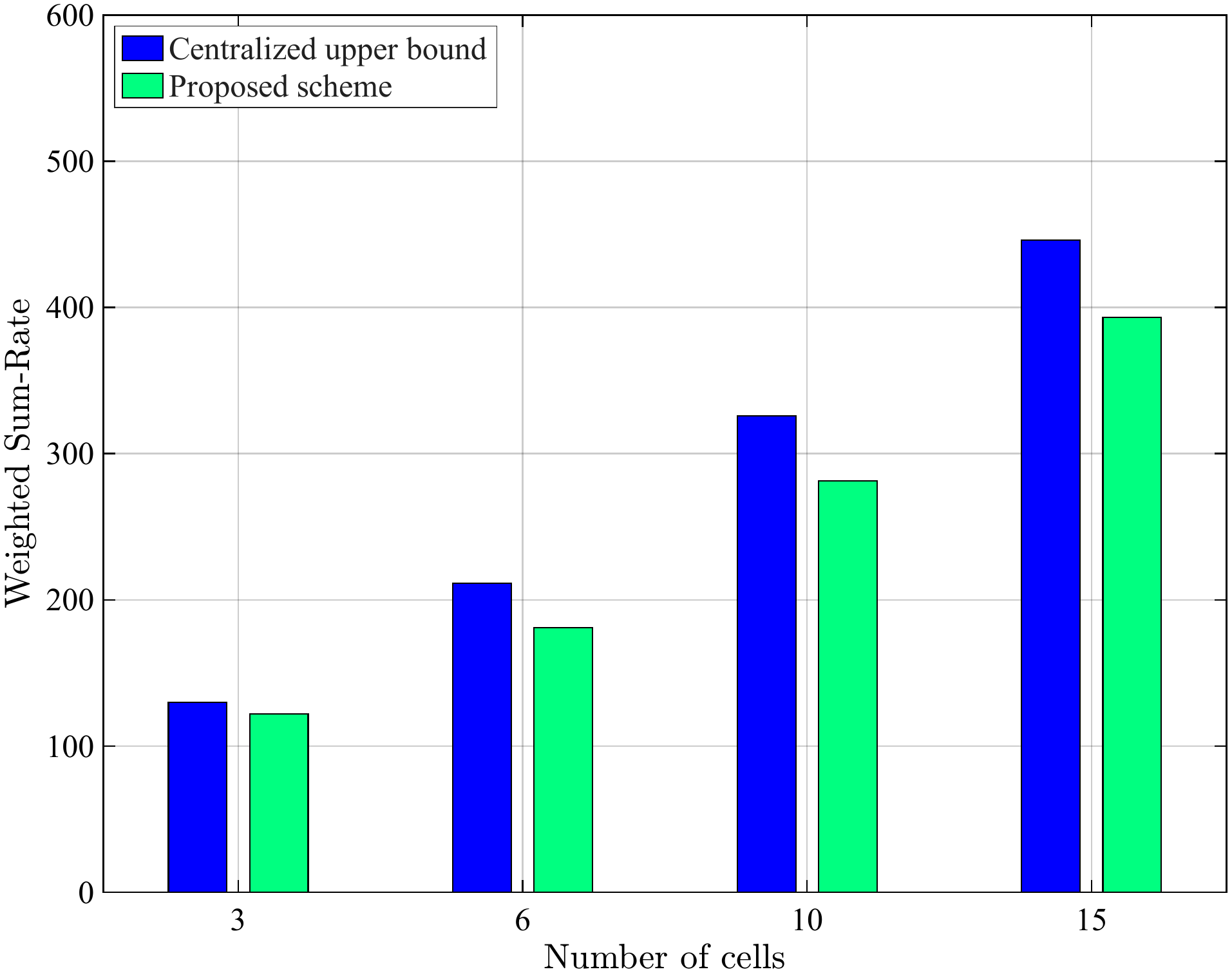}
\caption{Scalability performance versus the number of cells. The weighted sum-rate achieved by the proposed scheme is compared with the centralized upper bound for networks with $M=3$, $6$, $10$, and $15$ cells.}
	\label{fig:scalability}
\end{figure}

Fig.~\ref{fig:scalability} shows the network weighted sum-rate achieved by the proposed scheme and the centralized upper bound for different network sizes. As the number of cells increases from $M=3$ to $M=15$, the weighted sum-rates of both schemes increase accordingly, since more cells and users are simultaneously served in the network. More importantly, the proposed decentralized scheme consistently remains close to the centralized upper bound for all considered values of $M$, which demonstrates that the proposed IPC coordination and decentralized rotation optimization achieve performance that scales favorably with the network size. The overall trend indicates that the proposed method is still able to preserve most of the performance gain of the centralized design without requiring full network-wide optimization. These results confirm that the proposed distributed   two-timescale optimization framework maintains good effectiveness even when the BS adjacency graph becomes larger and the coordination task becomes more challenging.

\section{Conclusion}

In this paper, we investigated a multi-cell 6DMA-enabled downlink network under ICI and studied the joint design of short-term transmit precoding and long-term antenna rotations. By considering a practical cellular architecture with multiple 6DMA surfaces mounted on each BS, we formulated an average weighted sum-rate maximization problem under geometric feasibility constraints on the surface rotations. To address the strong inter-cell coupling and the inherent two-timescale nature of the problem, we proposed an IPC-assisted distributed framework, where the short-term precoders are optimized locally under coordinated IPC thresholds, while the IPC thresholds and 6DMA rotations are respectively updated through edge-wise inter-BS coordination and per-BS rotation optimization. Numerical results showed that the proposed distributed design achieves performance close to the centralized upper bound under different interference conditions and remains effective as the network size increases. These results suggest that the proposed framework provides an efficient and scalable approach for exploiting the spatial flexibility of 6DMA in practical multi-cell wireless networks.

\bibliographystyle{IEEEtran}
\bibliography{ref}

\end{document}